\documentclass[%
reprint,
superscriptaddress,
nofootinbib,
amsmath,
amssymb,
aps,
prd,
floatfix,
showkeys,
]{revtex4-2}

\usepackage{graphicx}
\usepackage{dcolumn}
\usepackage{bm}
\usepackage[normalem]{ulem}
\usepackage{fancyhdr}
\usepackage{graphicx,amsfonts,amssymb,amsbsy}
\usepackage{amsmath,amsthm,latexsym}
\usepackage[utf8]{inputenc}  
\usepackage{natbib}
\usepackage[colorlinks]{hyperref}
\usepackage{booktabs}
\usepackage{aas_macros}
\usepackage{xcolor}
\usepackage{xspace}
\usepackage{soul}
\usepackage{comment}

\hypersetup{
    citecolor=blue,  
    linkcolor=red,  
    urlcolor=blue    
}

\newcommand{\soft}{\textit{soft}\xspace}
\newcommand{\inter}{\textit{intermediate}\xspace}
\newcommand{\stiff}{\textit{stiff}\xspace}

\begin{document}

\title{Systematic study of the morphology and length of slow stable hybrid star branches}

\author{Mauro Mariani}
\email{mmariani@fcaglp.unlp.edu.ar}
\affiliation{Grupo de Astrofísica de Remanentes Compactos, Facultad de Ciencias Astronómicas y Geofísicas, Universidad Nacional de La Plata, Paseo del Bosque S/N, La Plata (1900), Argentina}
\affiliation{CONICET, Godoy Cruz 2290, Buenos Aires (1425), Argentina} 

\author{Milva G. Orsaria}
\email{morsaria@fcaglp.unlp.edu.ar}
\affiliation{Grupo de Astrofísica de Remanentes Compactos, Facultad de Ciencias Astronómicas y Geofísicas, Universidad Nacional de La Plata, Paseo del Bosque S/N, La Plata (1900), Argentina}
\affiliation{CONICET, Godoy Cruz 2290, Buenos Aires (1425), Argentina} 

\author{Germán Lugones}
\email{german.lugones@ufabc.edu.br}
\affiliation{Universidade Federal do ABC, Centro de Ciências Naturais e Humanas, Avenida dos Estados 5001- Bangú, CEP 09210-580, Santo André, SP, Brazil.}

\author{Ignacio F. Ranea-Sandoval}
\email{iranea@fcaglp.unlp.edu.ar}
\affiliation{Grupo de Astrofísica de Remanentes Compactos, Facultad de Ciencias Astronómicas y Geofísicas, Universidad Nacional de La Plata, Paseo del Bosque S/N, La Plata (1900), Argentina}
\affiliation{CONICET, Godoy Cruz 2290, Buenos Aires (1425), Argentina}

\begin{abstract}
We introduce and systematically study the length of the slow stable hybrid star branch as a quantitative measure of the extended stability region that arises in hybrid neutron stars when the hadron--quark phase conversion is slow compared to the radial oscillation timescale. Combining generalized piecewise-polytropic hadronic equations of state of varying stiffness with a constant-speed-of-sound quark-matter model, we construct a large set of hybrid equations of state spanning a broad range of transition pressures, energy-density jumps, and quark-matter speeds of sound. We identify four morphological types for the slow stable branch in the mass-radius plane: waterfall branches that descend monotonically from the hadronic maximum mass, bridges that connect the hadronic branch to a second unconditionally stable hybrid branch, tails that extend briefly beyond the maximum mass of an unconditionally stable hybrid branch, and tail-bridges that combine features of the latter two. Their prevalence is governed primarily by the transition pressure and the energy-density jump, while the branch length is also significantly influenced by the stiffness of the hadronic sector and the quark-matter speed of sound. Imposing current astrophysical and microphysical constraints shows that viable long branches are predominantly of waterfall type, and that stiff hadronic equations of state ---strongly disfavored under the rapid-conversion assumption--- remain compatible with all current constraints within the slow-conversion framework. In the plane of transition baryon density versus density jump, slow stable configurations open a new region of viable parameter space inaccessible under rapid conversions.
\end{abstract}

\maketitle

\section{Introduction}
\label{sec:intro}

Over the past decade, a wealth of astronomical data has placed increasingly stringent constraints on the equation of state (EoS) of dense matter. Precise mass measurements from radio pulsar timing have established that the maximum neutron star (NS) mass must exceed $\sim 2\,M_\odot$~\cite{Demorest:2010sdm, Antoniadis:2013amp, Cromartie:2020rsd, Fonseca:2021rfa}. The detection of gravitational waves from the binary NS merger GW170817~\cite{Abbott:2017gwa, Abbott:2018gmo} has provided complementary constraints on the tidal deformability, disfavoring very stiff EoSs. Meanwhile, X-ray observations by the NICER mission have yielded simultaneous mass-radius estimates for PSR~J0030+0451~\cite{Miller:2019pjm, Riley:2019anv} and PSR~J0740+6620~\cite{Miller:2021tro, Riley:2021anv}, further narrowing the allowed EoS space. More recent gravitational-wave~\cite{Abbott:2020goo} and X-ray~\cite{Salmi:2024anv, Choudhury2024anv, Mauviard:2025anv} observations have also established constraints on NS masses and radii. On the high-mass end, the analysis of gravitational-wave and kilonova data from GW170817 suggests an upper bound of $M_{\max} \lesssim 2.3\,M_\odot$~\cite{Rezzolla:2018ugw, Shibata:2019ctb, Musolino:2024otm}. At the opposite extreme, the mass and radius estimates for the central compact object in HESS~J1731$-$347, reported by \citet{Doroshenko:2022nwp} as $M = 0.77^{+0.20}_{-0.17}\,M_\odot$ and $R = 10.4^{+0.86}_{-0.78}$~km, suggest the existence of remarkably light and compact objects (though these estimates rely on specific assumptions about the stellar atmosphere and remain subject to debate~\cite{Alford:2023dcc}). Moreover, the mass and radius of the object XTE~J1814$-$338 have been estimated to be \mbox{$M = 1.21 \pm 0.05\,M_\odot$} and $R = 7.0 \pm 0.4$~km~\cite{kini:2024ctp}. As in the case of HESS~J1731$-$347, the validity of these estimates is still under discussion.

Relevant constraints also arise from nuclear and particle physics data. The properties of matter at densities $n \lesssim n_0$, where $n_0 = 0.16~\mathrm{fm}^{-3}$ is the nuclear saturation density, are well determined by nuclear theory and experiment. In the range immediately above, $n_0 \lesssim n \lesssim 2\,n_0$, several chiral effective field theory ($\chi$EFT) calculations provide constraints~\cite{Hebeler:2013nza, Lynn:2016ctn, Hu:2017nmp, Holt:2017eos, Drischler:2020hwd}; in this work, we implement the results of \citet{Drischler:2021lma}, who obtained a refined constraint for $\beta$-stable NS matter up to $2\,n_0$ using a Bayesian framework. On the high-density end, $n \gtrsim 40\,n_0$, perturbative Quantum Chromodynamics (pQCD) calculations yield constraints on the EoS of deconfined matter~\cite{Annala:2020efq, Gorda:2018gpy}. Although such extreme densities are typically absent in hadronic NS cores, the pQCD constraint can prove relevant in more exotic scenarios, as we will discuss.

Satisfying all of these constraints simultaneously is a nontrivial challenge for any single EoS model. In particular, a hadronic EoS stiff enough to support $\sim 2\,M_\odot$ stars typically predicts large radii and tidal deformabilities that are in tension with GW170817, and may fail to accommodate low-mass, compact objects such as the one reported in HESS~J1731$-$347. Conversely, a soft EoS that naturally explains compact configurations struggles to reach the $2\,M_\odot$ threshold once a phase transition is introduced. These tensions have motivated the study of hybrid stars (HSs), in which a hadronic mantle surrounds an inner core of deconfined quark matter, connected through a first-order phase transition~\cite{Laskos:2024hsi, Sagun:2023wit, Mariani:2022omh}.

A key aspect of this problem is the nature of the phase conversion at the hadron-quark interface when the star undergoes radial perturbations. The classical theorems on the radial stability of compact stars, dating back to the 1960s~\cite{harrison1965gravitation}, establish that the fundamental radial oscillation mode becomes unstable at the maximum mass configuration, where $\partial M / \partial \varepsilon_c = 0$, rendering all stellar models beyond this point (i.e., those with $\partial M / \partial \varepsilon_c < 0$) dynamically unstable. However, these results implicitly assume that matter is in full chemical (beta) equilibrium at all times ---i.e., that the reactions restoring equilibrium proceed instantaneously compared to the oscillation period (the so-called \emph{rapid} reaction regime). This assumption may break down at the hadron-quark interface inside a HS. If the reactions governing the phase conversion at the interface are \emph{slow} compared to the oscillation timescale, the interface oscillates in phase with the perturbation without achieving chemical equilibrium at each cycle. Under these conditions, \citet*{Pereira:2017rmp} demonstrated that HS configurations beyond the maximum mass ---where $\partial M / \partial \varepsilon_c < 0$--- can remain dynamically stable, since the fundamental mode frequency remains real and positive. These configurations constitute the slow stable HS (SSHS) branch: an extended sequence of stable HSs that exists only when the phase conversion is slow, connecting continuously to the standard stable branch but extending to lower masses and smaller radii beyond the turning point. The existence and properties of SSHS branches have been explored in numerous works~\cite{Pereira:2017rmp, mariani:2019mhs, Ranea:2022bou, Ranea:2023cmr, Ranea:2023auq, Rau:2023tfo, lugones:2023ama, Gosh:2024ero, Rather:2024roo, Celi:2025tau}, and it has been shown that they can significantly extend the range of masses and radii accessible to HSs, potentially resolving tensions between different observational constraints.

Previous studies have demonstrated that SSHSs can account for specific observations, such as the low-mass compact object in HESS~J1731$-$347~\cite{Mariani:2024cas} or XTE J1814-338~\cite{Mariani:2026hns}, and that ultra-stiff hadronic EoSs ---which would otherwise violate constraints from GW170817 or the maximum mass upper bound--- can become viable when the softening from the quark phase and the extended stability provided by slow conversion are taken into account~\cite{lugones:2023ama}. However, a systematic study of the SSHS branch itself, and in particular of its morphology and extent in the $M$-$R$ plane as a function of the EoS parameters, has not been carried out so far.

In this work, we address this gap by introducing and studying the \emph{length} of the SSHS branch as a quantitative measure of the extent of the extended stability region, as well as providing a qualitative characterization of its morphology. We construct a large set of hybrid EoSs using parametric models for both the hadronic and quark phases, and systematically map the SSHS branch length across the parameter space of transition pressure, energy density jump, and speed of sound in the quark phase. Our analysis reveals how the interplay between these parameters determines whether the SSHS branch is capable of satisfying diverse astrophysical constraints, and identifies the regions of parameter space that are most favorable for producing suitable astrophysical extended stable branches. This provides a unified framework to assess the viability of the slow conversion scenario and its observational implications.

The paper is organized as follows. In Sec.~\ref{sec:stability} we review the stability criteria for HSs with slow phase conversions, introduce the definition of the SSHS branch length, and discuss the applicability of the Seidov limit. In Sec.~\ref{sec:eos} we describe the parametric models used to construct the hybrid EoSs and the exploration of the parameter space that leads to the EoS sample analyzed in this work. Sections~\ref{sec:results1_len} and~\ref{sec:results2_astro} present the main results of our study, and Sec.~\ref{sec:conclus} summarizes our findings and discusses their astrophysical implications.

\section{Stability of hybrid stars with slow phase conversions and the SSHS branch length}
\label{sec:stability}

In this section, we review the radial stability theory relevant to HSs with a sharp hadron-quark interface, introduce the key concepts and terminology used throughout the paper, and define the SSHS branch length as a quantitative measure of the extended stability region.

\subsection{Radial stability and the turning-point criterion}
\label{sec:radial_stability}

The dynamical stability of a compact star against radial perturbations is determined by the spectrum of its radial oscillation modes. Assuming a harmonic (\textit{i.e.} $\exp (i \omega t)$) time dependence for the perturbations, this spectrum is governed by a Sturm-Liouville problem with ordered eigenvalues $\omega_n^2$. A stellar configuration is dynamically stable if the squared frequency of its lowest frequency mode, the fundamental one, satisfies $\omega_0^2 > 0$, and becomes unstable when $\omega_0^2 < 0$, {which renders an exponentially growing behavior for that particular mode indicating its unstable nature. The boundary between stable and unstable configurations is therefore defined by the condition $\omega_0^2 = 0$.

For a one-parameter family of stellar models constructed from a single-phase EoS ---such as purely hadronic stars composed of cold, catalyzed matter--- the classical results of Harrison, Thorne, Wakano, and Wheeler~\cite{harrison1965gravitation} establish that this stability boundary coincides with the maximum mass configuration along the sequence parametrized by the central energy density $\varepsilon_c$. Specifically, the fundamental mode becomes unstable precisely at the turning point where $\partial M / \partial \varepsilon_c = 0$, and all configurations with $\partial M / \partial \varepsilon_c < 0$ are dynamically unstable. This turning-point criterion provides a simple and powerful tool: one need only locate the maximum of $M(\varepsilon_c)$ to identify the last stable configuration, without solving the full oscillation eigenvalue problem.

\subsection{Phase conversion at the hadron-quark interface}
\label{sec:phase_conversion}

The situation becomes more complex in HSs with a sharp first-order phase transition between hadronic and quark matter. In this scenario, the EoS is discontinuous at the hadron-quark interface: the energy density jumps by $\Delta\varepsilon$ at constant pressure $P_t$. This discontinuity introduces a boundary condition in the radial oscillation eigenvalue problem that is absent in single-phase stars. Crucially, the form of this boundary condition depends on the \emph{speed} of the reactions governing the hadron-quark phase conversion relative to the oscillation timescale.

\paragraph*{Rapid conversion regime.}
If the reactions that convert hadronic matter into quark matter (and vice versa) proceed much faster than the oscillation period, the interface adjusts instantaneously to maintain chemical (beta) equilibrium at all times. In this regime, matter flows freely across the interface in response to radial perturbations, and the appropriate junction condition at the discontinuity allows for mass transfer between the two phases. Under these conditions, the standard turning-point criterion is recovered: the last stable configuration occurs at $\partial M / \partial \varepsilon_c = 0$, and all configurations with $\partial M / \partial \varepsilon_c < 0$ are dynamically unstable, just as in the single-phase case.

\paragraph*{Slow conversion regime.}
If instead the phase conversion reactions are slow compared to the oscillation timescale, the interface cannot equilibrate during each oscillation cycle. Matter on either side of the discontinuity oscillates in phase with the perturbation, but no net conversion occurs: the hadronic and quark phases behave effectively as immiscible fluids on the timescale of the perturbation. As demonstrated by \citet*{Pereira:2017rmp}, the appropriate junction conditions in this regime require that both the Lagrangian displacement $\xi$ and the Lagrangian perturbation of the pressure $\Delta P$ be continuous across the interface, in contrast to the rapid case where continuity of $\Delta P$ alone suffices.

These modified boundary conditions have a profound consequence: they alter the eigenvalue spectrum of the radial oscillation modes in a way that can keep the fundamental mode frequency real and positive ($\omega_0^2 > 0$) for configurations beyond local maxima, where $\partial M / \partial \varepsilon_c < 0$. The turning-point criterion therefore \emph{does not apply} in the slow conversion scenario: destabilization does not occur necessarily at the maximum mass star, but at the configuration at which $\omega_0^2$ eventually vanishes further along the hybrid sequence, at higher central densities. 

Section 3.4 of Ref.~\cite{mariani:2019mhs} provides a clear explanation of how to identify stellar configurations with $\omega_0^2=0$ directly from TOV equations, without solving the full radial perturbation problem\footnote{A detailed description of the same strategy is presented under the name of \textit{Residual Method} in Ref.~\cite{Zhang:2026cas-arXiv}.}.

The question of which regime ---rapid or slow--- is physically realized at the hadron-quark interface remains an open problem in QCD, as it depends on the poorly known kinetics of the deconfinement transition under the extreme conditions prevailing in NS cores. Both scenarios are therefore of theoretical and astrophysical interest, and a complete stability analysis of HSs must consider both possibilities.

\subsection{Unconditionally stable configurations and the SSHS branch}
\label{sec:uncond_stable}

The existence of two distinct stability regimes motivates a terminological distinction that will be used throughout this work. As discussed above, there are configurations that remain stable against radial perturbations regardless of the conversion speed at the interface. These \textit{unconditionally stable} (US) configurations satisfy $\partial M / \partial \varepsilon_c > 0$ and possess $\omega_0^2 > 0$ under both sets of junction conditions \cite{Pereira:2017rmp}\footnote{These configurations have been referred to as ``totally stable'' or ``fully stable'' in previous works~\cite{ lugones:2023ama, Mariani:2024cas, Mariani:2026hns}.}.

In contrast, configurations with $\partial M / \partial \varepsilon_c < 0$ that are stable ($\omega_0^2 > 0$) only in the slow conversion framework ---but unstable ($\omega_0^2 < 0$) under the rapid junction conditions--- constitute the SSHS branch~\cite{lugones:2023ama}. This branch begins contiguously after the last US configuration (which, in the most common case, coincides with the maximum mass star) and extends toward higher central densities, lower masses, and smaller radii. The SSHS branch terminates either at a stellar model for which $\omega_0^2$ vanishes under the slow junction conditions, or at the configuration where unconditional stability is recovered.

It has been shown in numerous works~\cite{Pereira:2017rmp, mariani:2019mhs, Ranea:2022bou, Ranea:2023cmr, Ranea:2023auq, Rau:2023tfo, lugones:2023ama, Gosh:2024ero, Rather:2024roo} that the SSHS branch can significantly extend the range of masses and radii accessible to HSs, potentially resolving tensions between different observational constraints

\subsection{Definition of the SSHS branch length}
\label{sec:branch_length}

The extent of the SSHS branch in the $M$-$R$ plane varies enormously depending on the EoS parameters, ranging from negligibly short sequences to branches that span more than a solar mass in gravitational mass and several kilometers in radius. In order to characterize this extent quantitatively, we introduce the \emph{length} of the SSHS branch, $L$, defined as
\begin{equation} 
\label{eq:longitud}
    L = \int_{{\rm SSHS}} \sqrt{{\rm d}\hat{M}^2 + {\rm d}\hat{R}^2} \, ,
\end{equation}
where $\hat{M}=M/(1.4\,M_\odot)$ and $\hat{R}=R/(12\,\rm{km})$. The normalization constants $1.4\,M_\odot$ and $12$~km are chosen so that $L$ is a dimensionless quantity and so that the mass and radius contributions enter the sum on comparable footing: with this choice, a variation of $1.4\,M_\odot$ in mass and $12$~km in radius ---representative of the typical scales of NS properties--- contribute equally to the total length, preventing either term from dominating. The integral of Eq.~\eqref{eq:longitud} is evaluated along the sequence of stable SSHS configurations parameterized by the central energy density $\varepsilon_c$, starting from the last US configuration (usually the maximum-mass star of the hadronic branch) and ending at the configuration where $\omega_0^2=0$ under slow conversion conditions\footnote{The terminal configuration is identified by integrating the radial pulsation equations jointly with the TOV structure equations setting $\omega_0^2=0$, and locating the stellar model along the hybrid sequence at which the surface boundary condition $\Delta P(R)=0$ is satisfied. This shooting-type strategy follows our previous works~\cite{mariani:2019mhs, Ranea:2022bou,lugones:2023ama, Mariani:2024cas, Mariani:2026hns} and avoids solving the full eigenvalue problem at every stellar model.} (or at the configuration where unconditional stability is recovered). 

It should be noted that, in addition to the primary SSHS branch ---defined as the first slow-stable segment encountered along the stellar sequence with increasing central density---, secondary slow-stable segments may appear at higher central densities, separated from the primary branch by an intervening US hybrid branch (see the discussion in Sec.~\ref{sec:morphology}).  These secondary segments are not included in the computation of $L$, as the EoS parametrizations that produce them typically fail the $M_{\max} \geq 2.01\,M_\odot$ constraint and/or exhibit lengths that are negligible compared to that of the primary branch; their systematic characterization is deferred to a future study. Accordingly, $L$ quantifies the extent of the primary SSHS branch in the normalized $M$-$R$ plane, providing a single scalar measure that enables a systematic comparison across the full parameter space of hybrid EoSs.

It is worth noting that the branch length $L$, as defined by Eq.~\eqref{eq:longitud}, is a geometric quantity that measures the arc length traversed in the normalized $M$-$R$ plane. It does not directly encode the number of stable configurations, the range of central densities spanned, or the astrophysical viability of the branch. These properties will be analyzed separately in Secs.~\ref{sec:results1_len} and~\ref{sec:results2_astro}. Nevertheless, as we will show, $L$ proves to be a useful and intuitive measure that correlates well with the astrophysical relevance of the SSHS branch, as longer branches generically reach more extreme regions of the $M$-$R$ plane.

\subsection{The Seidov condition}
\label{sec:seidov}

A classical result due to \citet{Seidov:1971tso} establishes a criterion for determining the sign of $\partial M / \partial \varepsilon_c$ at the point where the new phase first appears. This result relates the value of $\partial M / \partial \varepsilon_c$ at the transition onset to the strength of the first-order phase transition, quantified by the energy density discontinuity $\Delta\varepsilon$ at the transition pressure $P_t$. Specifically, $\partial M / \partial \varepsilon_c < 0$ at the quark core appearance if the energy density jump across the interface satisfies $\Delta\varepsilon \geq \Delta\varepsilon_{\rm crit}$, where the critical value is
\begin{equation}
\Delta\varepsilon_{\rm crit} = \tfrac{1}{2}\,\varepsilon_{H,t} + \tfrac{3}{2}\,P_t \,,
\label{eq:seidov_delta}
\end{equation}
with $\varepsilon_{H,t}$ the energy density on the hadronic side of the interface. When $\Delta\varepsilon > \Delta\varepsilon_{\rm crit}$, the onset of the quark core triggers a turning point in $M(\varepsilon_c)$; when $\Delta\varepsilon < \Delta\varepsilon_{\rm crit}$, it does not.

Within the rapid conversion scenario, configurations with $\partial M / \partial \varepsilon_c < 0$ are dynamically unstable, so the Seidov limit effectively determines whether the phase transition destabilizes the star at the onset of the quark core. In this work, however, we adopt a different perspective: since in the slow conversion framework configurations with $\partial M / \partial \varepsilon_c < 0$ can remain stable (forming the SSHS branch), the Seidov condition does not act as a stability boundary but rather as a tool for classifying the morphology of the $M$-$R$ curves. Whether $\Delta\varepsilon$ lies above or below $\Delta\varepsilon_{\rm crit}$ determines whether the SSHS branch appears immediately at the onset of the quark core or only further along the hybrid sequence, beyond a segment of US hybrid configurations.

It should be emphasized, however, that the Seidov condition is a \emph{local} criterion: it determines the sign of $\partial M / \partial \varepsilon_c$ at the onset of the phase transition, but it does not predict the global structure of the stellar sequence at higher central densities. In particular, configurations with $\Delta\varepsilon$ above $\Delta\varepsilon_{\rm crit}$ may still recover unconditional stability at higher densities if the post-transition EoS is sufficiently stiff, while configurations below the limit may develop SSHS segments further along the sequence due to the cumulative softening from the quark core. 

In order to apply the Seidov condition to our analysis in the $M$-$R$ plane, it is important to recall that the zeros of the $\partial M / \partial \varepsilon_c$ derivative coincide with those of the $\partial M / \partial R$ derivative. The full morphological classification and its dependence on the interplay of the hadronic and quark EoS properties are presented in Sec.~\ref{sec:morphology}.

\section{Equation of state model}
\label{sec:eos}

In this work, we model HSs composed of a hadronic mantle and a quark matter inner core, connected through an abrupt first-order phase transition. Rather than relying on a specific microscopic model for each phase, we employ parametric descriptions for both the hadronic and quark sectors. This approach allows us to systematically explore a broad region of the hybrid EoS parameter space and to draw general, model-independent conclusions regarding the structure and properties of the SSHS branch.

\begin{table*}[tb]
\caption{Parameters and characteristic stellar properties of the three hadronic EoSs constructed with the GPP prescription by \citet{OBoyle:2020peo}. In all cases, $\log_{10}(\rho_0) = 14.127$ and $\log_{10}(K_1) = -27.22$ (so that the hadron outer core begins at $0.5\,n_0$), with densities in $\mathrm{g\,cm^{-3}}$. The three EoSs are consistent with the $\chi$EFT band of Ref.~\cite{Drischler:2021lma} at low densities. The stellar properties listed are: radius of the canonical $M = 1.4\,M_\odot$ configuration ($R_{1.4}$), mass and radius of the maximum mass configuration ($M_{\max}$ and $R_{M_{\max}}$), and central baryon number density and pressure at the maximum mass configuration ($n_{B,c}^{M_{\max}}$ and $P_c^{M_{\max}}$).}
\centering
\begin{ruledtabular}
\begin{tabular}{lcccccccccc}
 & $\log_{10}\rho_1$ & $\log_{10}\rho_2$ & $\Gamma_1$ & $\Gamma_2$ & $\Gamma_3$ 
 & $R_{1.4}$ & $M_{\max}$ & $R_{M_{\max}}$ & $n_{B,c}^{M_{\max}}$ & $P_c^{M_{\max}}$ \\
 & & & & & 
 & [km] & $[M_\odot]$ & [km] & $[n_0]$ & [MeV/fm$^3$] \\
\hline
\soft  & 14.30 & 14.65 & 2.757 & 3.4 & 2.70 & 12.02 & 2.06 & 10.26 & 7.22 & 743.9 \\
\inter & 14.50 & 14.65 & 2.763 & 5.1 & 2.35 & 12.71 & 2.22 & 11.30 & 5.99 & 528.67 \\
\stiff & 14.55 & 14.65 & 2.766 & 9.5 & 1.05 & 13.24 & 2.47 & 13.15 & 4.26 & 254.30 \\
\end{tabular}
\end{ruledtabular}
\label{tabla:eos_had}
\end{table*}

\subsection{Hadronic phase}
\label{sec:hadroneos}

For the outer layers of the star, we adopt the BPS-BBP crust~\cite{Baym:1971tgs, Baym:1971nsm} up to $0.5\,n_0$. For the hadronic core at higher densities, we employ the generalized piecewise polytropic (GPP) parametrization introduced by \citet{OBoyle:2020peo}. In this framework, the hadronic core EoS is divided into three density segments delimited by the boundaries $\rho_0$, $\rho_1$, and $\rho_2$. The first boundary, $\rho_0$, marks the crust-core transition at $0.5\,n_0$, while $\rho_1$ and $\rho_2$ are mathematical boundaries between successive polytropic segments whose values are tuned to control the overall stiffness of the EoS while preserving continuity across segments.

Within each segment $[\rho_{i-1},\,\rho_i]$, the rest-mass density $\rho$, energy density $\varepsilon$, and speed of sound $c_s$ are given as functions of the pressure $P$ by~\cite{OBoyle:2020peo}
\begin{align}
\rho &= \left(\frac{P - \Lambda_i}{K_i}\right)^{1/\Gamma_i} \, , \label{eq:gpp_rho} \\[6pt]
\varepsilon &= \frac{K_i\,\rho^{\,\Gamma_i}}{\Gamma_i - 1} + (1 + a_i)\,\rho - \Lambda_i \, , \label{eq:gpp_eps} \\[6pt]
c_s &= \left[\frac{1}{\Gamma_i - 1} + \frac{1 + a_i}{K_i\,\Gamma_i\,\rho^{\,\Gamma_i - 1}}\right]^{-1/2} \, , \label{eq:gpp_cs}
\end{align}
where $\Gamma_i$ is the polytropic index and $K_i$ is the polytropic constant for each segment. The auxiliary parameters $a_i$ and $\Lambda_i$ are determined by requiring continuity and differentiability of $\varepsilon(\rho)$ and $P(\rho)$ at each dividing density, which ensures that $c_s$ is also continuous across segment boundaries. Following Refs.~\cite{OBoyle:2020peo, Read:2008iy}, we incorporate the speed of light $c$ into the definitions of the energy density and pressure, so that $\rho$, $\varepsilon$, and $P$ are expressed in $\mathrm{g\,cm^{-3}}$.

The flexibility of this parametrization allows us to construct EoSs with a wide range of stiffness while anchoring them to well-established low-density nuclear physics. Specifically, the values of $\log_{10}\rho_0$ and $\log_{10}K_1$ are chosen so that the crust-core interface occurs at $0.5\,n_0$ and the first polytropic segment matches the pressure and energy density predicted by chiral effective field theory ($\chi$EFT) calculations~\cite{Drischler:2021lma}, serving as a low-density constraint for our phenomenological core EoS rather than implying the direct use of a $\chi$EFT model.

Using this parametrization, we construct three representative hadronic EoSs ---\soft, \inter, and \stiff--- whose parameters and characteristic stellar properties are listed in Table~\ref{tabla:eos_had}, and whose pressure-density relations are shown in Fig.~\ref{fig:eosmr_all}(a). These same parametrizations were implemented previously in Ref.~\cite{Mariani:2024cas}, with the focus on providing an alternative explanation for the low-mass ultra-compact star in the supernova remnant HESS~J1731$-$347. As can be seen in the figure, all three EoSs traverse the $\chi$EFT band of Ref.~\cite{Drischler:2021lma} at low densities and then separate at higher densities according to their stiffness. Through TOV integration, we obtain the corresponding stellar configurations and present them in the mass-radius plane of Fig.~\ref{fig:eosmr_all}(b), showing each sequence up to its maximum mass configuration. The \soft EoS produces the most compact stellar configurations, with radii around $\sim 12$~km and a maximum mass that only marginally exceeds $2\,M_\odot$. The \stiff EoS yields the largest radii ($\sim 14$~km) and the highest maximum mass ($\sim 2.4\,M_\odot$). As expected, the \inter EoS produces intermediate values of both quantities.

In the $M$-$R$ panel we also display the current astrophysical constraints, including $\sim 2\,M_\odot$ pulsars, gravitational-wave detections, and X-ray observations, as well as the causality-forbidden region in the upper-left corner and the upper bound $M_{\max} \lesssim 2.3\,M_\odot$ inferred from gravitational-wave and kilonova data~\cite{Rezzolla:2018ugw, Shibata:2019ctb, Musolino:2024otm}. Regarding these constraints, the \soft hadronic EoS satisfies all of them on its own, including the mass and radius estimates for HESS~J1731$-$347~\cite{Doroshenko:2022nwp}, which favor compact, low-mass configurations. The \inter EoS satisfies all constraints except HESS~J1731$-$347, as its larger radii at low masses prevent it from reaching this region. The \stiff EoS additionally does not meet the tidal deformability constraint of GW170817~\cite{Abbott:2017gwa,Abbott:2018gmo} and exceeds the upper bound $M_{\max} \lesssim 2.3\,M_\odot$~\citep{Shibata:2019ctb}. As in previous works~\citep{Mariani:2024cas, Mariani:2024csi}, these three hadronic EoS models are designed to span a broad range of stellar properties, guided by current microphysical and astrophysical constraints, and to serve as limiting representative cases that allow us to conduct our analysis within a generalized model-independent approach.

\begin{figure*}[t]
\centering
\includegraphics[height=.34\linewidth]{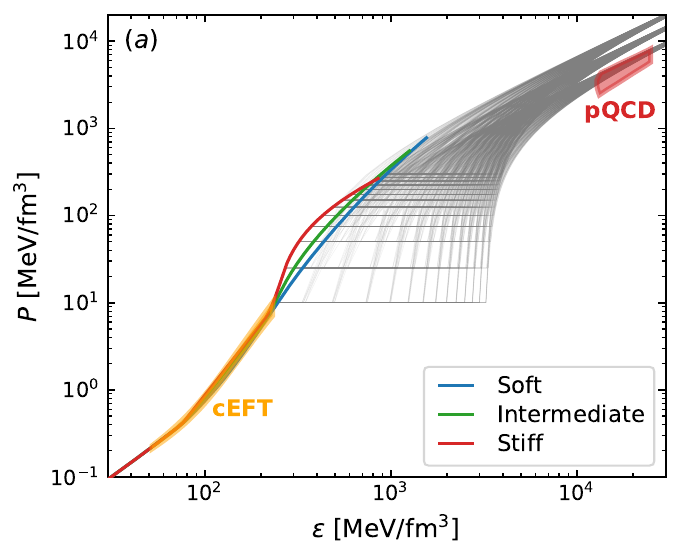}
\includegraphics[height=.34\linewidth]{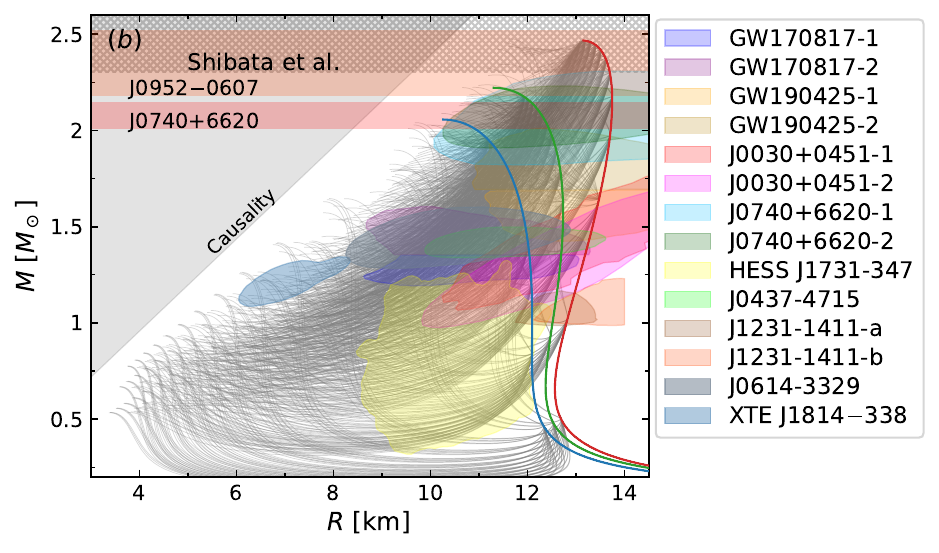}
\caption{(a) Pressure as a function of energy density for the three hadronic EoSs (\soft, \inter, and \stiff) and for all the hybrid EoSs constructed in this work (gray curves). The low-density $\chi$EFT~\cite{Drischler:2021lma} and high-density pQCD~\citep{Annala:2020efq} constraints are indicated by orange and red shaded regions, respectively. (b) Mass-radius relationships obtained by integrating the TOV equations for each EoS. Colored solid curves correspond to the three purely hadronic models; gray curves show the HS families. For each set, only stable configurations within the slow hadron-quark conversion scenario are shown. Colored regions and horizontal bands indicate current constraints (see text for details and references); the causality-forbidden zone is also displayed as a shaded gray region.}
\label{fig:eosmr_all}
\end{figure*}

\subsection{Quark phase and hybrid construction}
\label{sec:quarks}

For the high-density quark matter phase, we adopt the constant speed of sound (CSS) parametrization~\cite{Alford:2013aca}. The CSS model provides a minimal and general framework to represent qualitative features of various microscopic models of quark matter. It is characterized by three parameters: the hadron-quark transition pressure $P_t$, the energy density discontinuity at the transition $\Delta\varepsilon$, and the squared speed of sound in the quark phase $c_s^2$, which is taken to be constant and independent of pressure. This parametrization, combined with an abrupt first-order phase transition at $P = P_t$, connects the hadronic and quark branches through a discontinuity $\Delta\varepsilon$ in the energy density at constant pressure, as seen in the horizontal segments visible in Fig.~\ref{fig:eosmr_all}(a).

Combining the three hadronic EoSs of Sec.~\ref{sec:hadroneos} with the CSS parameters, we construct a total of
\begin{equation}
3\,(\text{hadron}) \times 13\,(P_t) \times 13\,(\Delta\varepsilon) \times 3\,(c_s^2) = 1521
\label{eq:total_eos}
\end{equation}
hybrid EoSs, spanning the following parameter ranges:
\begin{align}
10~\mathrm{MeV/fm^3}  &\leq P_t \leq 300~\mathrm{MeV/fm^3} \, , \nonumber \\
100~\mathrm{MeV/fm^3}  &\leq \Delta\varepsilon \leq 3000~\mathrm{MeV/fm^3} \, , \label{eq:css_ranges} \\
c_s^2 &= 0.33,\; 0.50,\; 0.70 \, . \nonumber
\end{align}
The chosen ranges are consistent with those employed in previous works based on microphysical models of hybrid EoSs. The $P_t$ values are in line with Refs.~\cite{ranea:2016css, ranea:2017csi, tonetto:2020dgm, Rather:2024roo}. The $\Delta\varepsilon$ range extends to values larger than those typically predicted by many microscopic calculations, but compatible with several other microscopic EoS models~\cite{mariani:2019mhs, Jarvinen:2022hmo, Mariani:2022omh, Lenzi:2023hsw, Celi:2025etr, Celi:2025tau}; retaining these large values is essential for exploring the full extent of the SSHS branch phenomenology. The values of $c_s^2$ span from the conformal limit ($c_s^2 = 1/3$) to stiffer quark matter consistent with non-perturbative QCD models (see, e.g., Refs.~\cite{Mclerran:2019qma, Gholami:2025aco}). Similar CSS parameter samplings were adopted in Refs.~\cite{lugones:2023ama, Mariani:2024cas, Mariani:2026hns}, where a limited set of EoSs was selected for detailed study; here, we instead perform a systematic analysis of the full sample in order to draw more general conclusions.

Within the selected range of $P_t$, we compute $L$ only for cases in which $P_t \leq P_c^{M_{\max}}$, i.e., cases where the phase transition occurs at or below the central pressure of the maximum mass hadronic configuration. For the \stiff EoS, this excludes  the narrow interval of transition pressures between $P_t \simeq 255$--$300~\mathrm{MeV/fm^3}$ for which $P_t > P_c^{M_{\max}}$ (see Table~\ref{tabla:eos_had}). The potential appearance of SSHS branches in this unconventional post-maximum-mass transition scenario has not been studied yet and falls outside the scope of the present work; we assign $L = 0$ to these cases and defer their analysis to a future study.

Figure~\ref{fig:eosmr_all} provides a global overview of the full sample of theoretical models considered in this work. Fig.~\ref{fig:eosmr_all}(a) shows the pressure-energy density relations, and Fig.~\ref{fig:eosmr_all}(b) the corresponding mass-radius curves obtained by TOV integration. In both cases, each hybrid EoS is composed of a hadronic branch (colored curve) up to the transition pressure, beyond which the quark sector appears (gray curve). As discussed in Sec.~\ref{sec:hadroneos}, all hybrid EoSs satisfy the low-density $\chi$EFT constraint by construction. Regarding the high-density pQCD constraint~\cite{Annala:2020efq}, only the EoSs with $c_s^2 = 0.33$ fulfill it across all densities, while those with $c_s^2 = 0.50$ and $0.70$ may violate the pQCD bound at sufficiently high densities ---see the light gray curves that miss the pQCD region in the upper-right portion of Fig.~\ref{fig:eosmr_all}(a). However, these EoSs are not discarded \textit{a priori}: they remain valid provided that the central density of the last stable stellar configuration does not exceed the pQCD threshold~\cite{Mariani:2024cas}.

In the $M$-$R$ diagram presented in Fig.~\ref{fig:eosmr_all}(b), all configurations along the presented curves are dynamically stable within the slow conversion scenario. The US configurations ---stable in both the rapid and slow scenarios--- correspond to the subset satisfying $\partial M / \partial \varepsilon_c > 0$. The resulting hybrid $M$-$R$ curves display a rich variety of morphologies depending on the CSS parameters. The presence of the phase transition generically softens the EoS, reducing the maximum mass and shifting stellar configurations toward smaller radii with respect to the purely hadronic case. The detailed analysis of these morphologies and their connection with the SSHS branch length is the subject of Sec.~\ref{sec:results1_len}.

\section{SSHS branch morphology and length}
\label{sec:results1_len}

In this section, we analyze in detail the sample of 1521 hybrid EoSs presented in Fig.~\ref{fig:eosmr_all}, focusing on the morphology and length of the SSHS branch. We first identify and classify the distinct morphological types that arise in the $M$-$R$ plane (Sec.~\ref{sec:morphology}), and then we perform a systematic study on the properties and parameter dependence of the branch length $L$ (Secs.~\ref{sec:impact_had} and \ref{sec:branch_diversity}).

\subsection{Morphological types and the role of the EoS parameters}
\label{sec:morphology}

We begin by identifying the main morphological types of SSHS branches that arise in the $M$-$R$ plane and the physical mechanisms that govern them. To this end, we first analyze schematic cases that isolate the key behaviors for representative parameter choices. After this, we will turn to the full sample of 1521 hybrid EoSs in Secs.~\ref{sec:impact_had} and~\ref{sec:branch_diversity}.

A related classification of HS topologies within the rapid conversion scenario, guided by the Seidov limit, was presented by \citet{Alford:2013aca} [cf.\ their Fig.~3]. Recalling the Seidov condition discussion, the phase transition strength (quantified by $\Delta \varepsilon$) determines the sign of $\partial M / \partial R$ at the stellar configuration in which the quark core first appears. The classification introduced here, formulated within the slow conversion framework, provides a complementary morphological perspective in the $M$-$R$ plane with a fundamentally different stability interpretation of the resulting branches.

\begin{figure}[tb]
\centering
\includegraphics[width=0.95\linewidth]{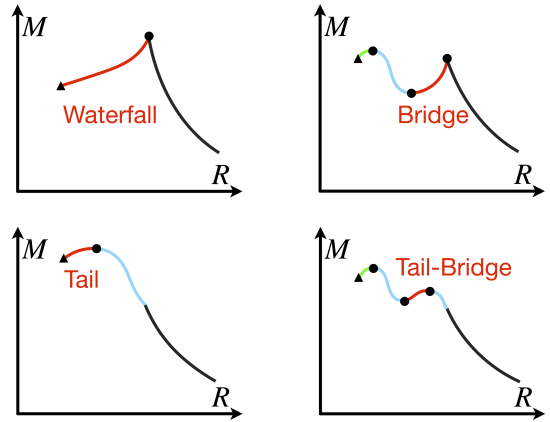}
\caption{Schematic mass-radius diagrams illustrating the four morphological types identified in this work. In each panel, the black curve denotes the stable hadronic branch, the light blue segments correspond to US hybrid configurations, and the red (green) segments indicate the primary (secondary) SSHS branch, stable only under the slow phase conversion hypothesis. Filled circles mark local maxima and minima (turning points), while filled triangles mark terminal configurations where $\omega_0^2 = 0$ (under slow reactions) without a turning point. See text for a detailed description of each morphology.}
\label{fig:sketch}
\end{figure}

\begin{figure*}[t]
\centering
\includegraphics[width=0.7\linewidth]{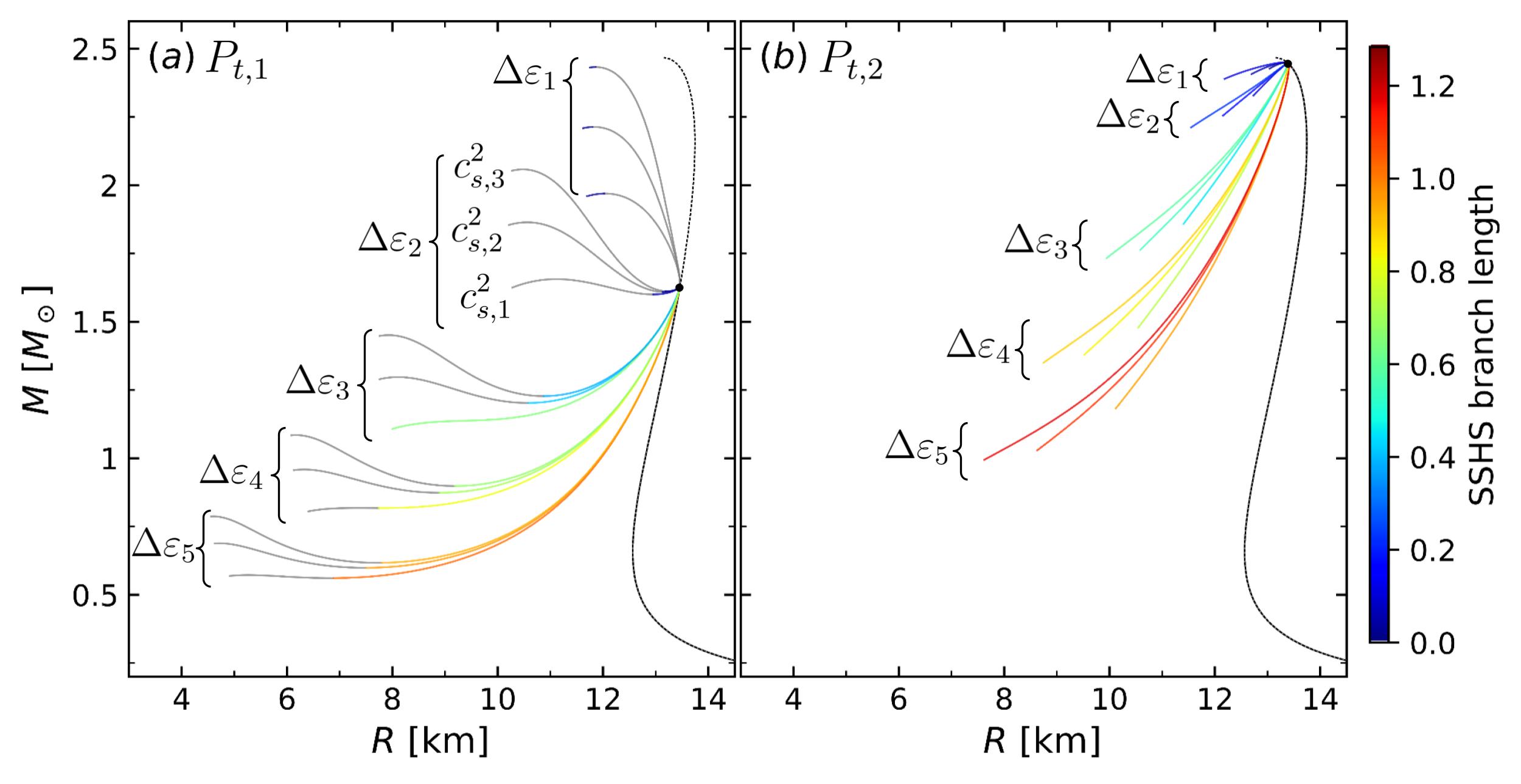}
\caption{Schematic mass-radius relationships illustrating the role of the CSS parameters on the SSHS branch morphology, for a low transition pressure $P_{t,1}$~(a) and a high transition pressure $P_{t,2}$~(b). In both cases, $\Delta\varepsilon_i < \Delta\varepsilon_{i+1}$; within each group of curves at fixed $\Delta\varepsilon$, the ordering $c^2_{s,1} < c^2_{s,2} < c^2_{s,3}$ holds, as explicitly labeled for $\Delta\varepsilon_2$ in~(a). Black curves show the purely hadronic sequences and gray segments denote US hybrid configurations; the primary SSHS branch is colored by $L$. The black dot marks the maximum mass of the purely hadronic sequence. See Sec.~\ref{sec:morphology} for details.}
\label{fig:mraio_selec}
\end{figure*}

Four distinct morphological types arise in our sample, distinguished by the role of the SSHS branch within the stellar sequence. Each type is defined by the structure of the stable configurations that appear beyond the onset of the quark core, and is illustrated schematically in Fig.~\ref{fig:sketch}. The classification refers exclusively to the primary SSHS branch (red curves in Fig.~\ref{fig:sketch}); secondary branches (green curves) are not analyzed in this work, as discussed in Sec.~\ref{sec:branch_length}. Through this figure it is possible to graphically establish the effects of the Seidov condition in the $M$-$R$ curve morphology. The onset of the quark core is located at the endpoint of the black curve in each panel; if $\Delta \varepsilon$ is below (above) the Seidov limit, $\Delta\varepsilon_\textrm{crit}$, the curve increases (decreases) after that point, indicated with a light-blue (red) segment. Considering this approach, the four types are:
\begin{itemize}
\item[\textbf{W}] \textbf{(Waterfall).} The SSHS branch originates at the hadronic turning point ---which coincides with the maximum mass of the sequence--- and extends monotonically toward lower masses and smaller radii, with no subsequent recovery of unconditional stability. This morphology prevails at high transition pressures and large values of $\Delta\varepsilon$ (hence the Seidov limit is crossed for this morphology), and is analogous to topology~$A$ of Ref.~\cite{Alford:2013aca}, although the stability interpretation differs owing to the distinct junction conditions.

\item[\textbf{B}] \textbf{(Bridge).} As in Type~W, the SSHS branch appears immediately after the onset of the quark core; however, it acts as an intermediate connection between the stable hadronic branch and a (hybrid) second US branch ---commonly referred to as the \textit{twin hybrid branch} or \textit{third family of compact stars}---. The length of this bridging segment varies widely depending on the EoS parameters. This morphology arises when $P_t$ is sufficiently low and $\Delta\varepsilon$ sufficiently large (the Seidov limit is also crossed for this morphology) for unconditional stability to be recovered beyond the phase transition, and corresponds to topology~$D$ of Ref.~\cite{Alford:2013aca}, extended here by the bridge connection inherent to the slow conversion scenario.

\item[\textbf{T}] \textbf{(Tail).} An US hybrid branch extends continuously from the hadronic branch up to the maximum mass of the sequence ---the standard hybrid scenario widely studied in the literature. Beyond this maximum, an SSHS tail appears, stable only under slow phase conversion, with no subsequent recovery of unconditional stability. Such tails are invariably short or extremely short. This morphology occurs for all $P_t$ values at sufficiently low $\Delta\varepsilon$ (the Seidov limit is not crossed for this morphology) and corresponds to topology~$C$ of Ref.~\cite{Alford:2013aca}, where no stable extension exists beyond the maximum-mass configuration in the rapid case.

\item[\textbf{TB}] \textbf{(Tail-Bridge).} Identical to Type~T (consequently, the Seidov limit is not crossed for this morphology), but with a subsequent recovery of unconditional stability through a second US hybrid branch. Related to topology~$B$ of Ref.~\cite{Alford:2013aca}, this is the least frequent morphology, as it requires fine-tuned parameter values. Because it produces very short initial US hybrid branches, it is not easily distinguishable from Type~B.
\end{itemize}

Morphologies W, B, T, and TB are specific to the slow conversion framework. In the rapid conversion scenario, only the standard hybrid branch ---US configurations contiguously connected to the hadronic branch via a phase transition below the Seidov limit--- and the disconnected second branch of USHSs can be obtained.

Having established this classification, we now examine how the model parameters govern the branch morphology. Independently of the hadronic EoS, we find that the morphologies identified above arise predominantly in two regimes, illustrated schematically in the two panels of Fig.~\ref{fig:mraio_selec}: an early quark onset at low $P_t$ produces morphologies that differ markedly from those obtained at high $P_t$, where the quark core appears late in the stellar sequence. We describe each regime in turn, focusing on the role of the quark EoS parameters.

\paragraph*{Early transitions:} Fig.~\ref{fig:mraio_selec}~(a) shows representative \mbox{$M$-$R$} curves for a relatively low transition pressure $P_{t,1}$, with increasing values of the energy density jump \mbox{($\Delta\varepsilon_1 < \Delta\varepsilon_2 < \cdots < \Delta\varepsilon_5$).} The black dot marks the maximum mass of the purely hadronic sequence. Gray segments correspond to US configurations; colored segments indicate the SSHS branch, with the color encoding $L$.

For the smallest density jump, $\Delta\varepsilon_1$, the system lies below the Seidov limit: the phase transition does not induce a turning point, and the hybrid sequence continues with $\partial M / \partial \varepsilon_c > 0$ beyond the onset of the quark core. The resulting morphology is Type~T: an US hybrid branch extends up to the maximum mass, followed by a very short SSHS tail (dark blue). The SSHS segments that appear in this regime are extremely short, as the fundamental mode becomes unstable almost immediately beyond the turning point. In a few extreme cases, the US hybrid branch is not followed by any SSHS segment, yielding $L = 0$.

For intermediate values of $\Delta\varepsilon$ between $\Delta\varepsilon_1$ and $\Delta\varepsilon_2$ (i.e., still below the Seidov limit), a Type~TB morphology may arise. This case is not shown in the figure, as it is relatively rare and, when present, the first US hybrid branch is too short to be clearly distinguishable from Type~B configurations.

\begin{figure*}[t]
\centering
\includegraphics[width=\linewidth]{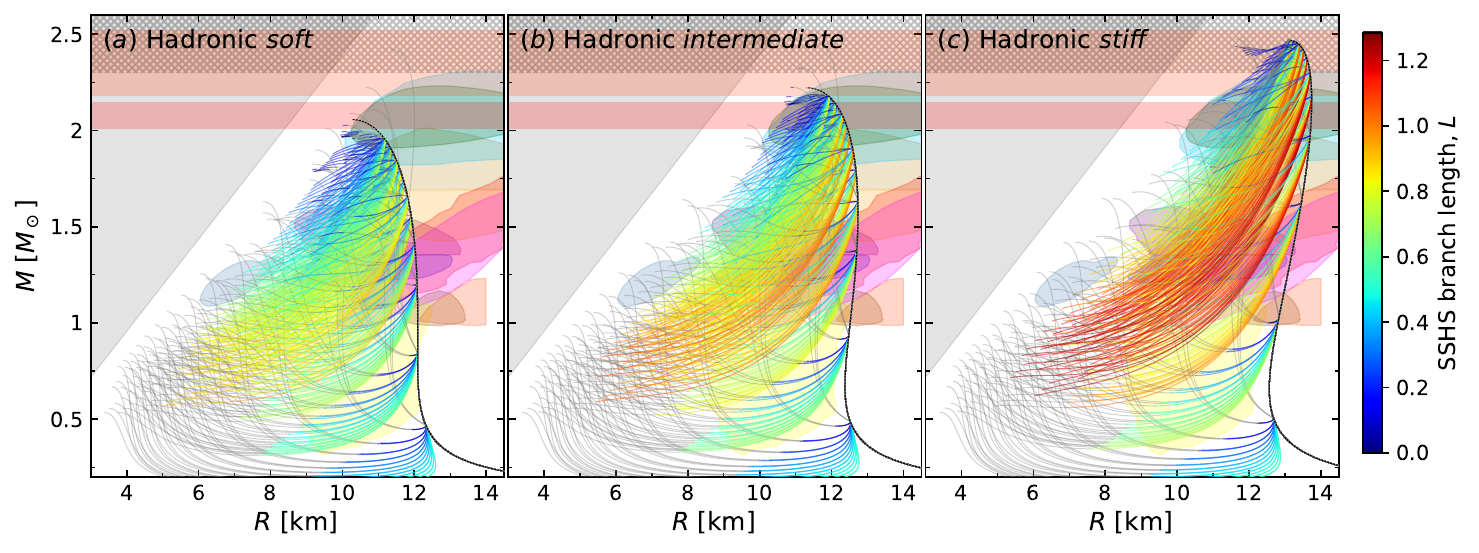}
\caption{Mass-radius relationships for all hybrid EoSs constructed in this work, separated by the underlying hadronic model: \soft~(a), \inter~(b), and \stiff~(c). Black dashed curves show the purely hadronic sequences; gray segments correspond to US hybrid configurations along each hybrid sequence. The SSHS branch is colored according to its length $L$ (color bar). Only dynamically stable configurations of the primary SSHS branch are shown. Observational constraints are the same as in Fig.~\ref{fig:eosmr_all}.}
\label{fig:mrlenghts_multi_had}
\end{figure*}

As $\Delta\varepsilon$ increases to $\Delta\varepsilon_2$, the system crosses the Seidov limit ---meaning that $\Delta\varepsilon$ exceeds the critical value $\Delta\varepsilon_{\rm crit}$ [Eq.~\eqref{eq:seidov_delta}] above which the phase transition destabilizes the star in the rapid conversion scenario. The morphology transitions to Type~B: the SSHS branch appears immediately after the onset of the quark core, connecting the stable hadronic branch to an US hybrid branch at lower masses. For $\Delta\varepsilon_2$, the three curves corresponding to different speeds of sound ($c^2_{s,1} < c^2_{s,2} < c^2_{s,3}$) illustrate the effect of $c_s^2$ on the length of the Type~B bridge: a \emph{lower} $c_s^2$ produces a \emph{longer} SSHS segment. The physical origin of this behavior is that a softer quark phase leads to a more pronounced softening of the post-transition EoS, which delays the recovery of unconditional stability and thereby extends the bridge. Conversely, a stiffer quark phase supports the hybrid sequence more effectively, allowing unconditional stability to be recovered sooner and truncating the SSHS segment (secondary SSHS segments that may appear beyond the US hybrid branch are not analyzed in this work; see Sec.~\ref{sec:branch_length}).

As $\Delta\varepsilon$ continues to increase ($\Delta\varepsilon_3 \to \Delta\varepsilon_5$), the Type~B bridges grow progressively longer (green $\to$ yellow $\to$ orange), while the US hybrid branches that follow tend to become shorter. For the largest values of $\Delta\varepsilon$, the SSHS bridge dominates the hybrid portion of the sequence, extending to low masses ($M \lesssim 1\,M_\odot$) and small radii ($R \sim 5$-$8$~km). 

\paragraph*{Late transitions:} Fig.~\ref{fig:mraio_selec}~(b) shows the contrasting behavior at a relatively high transition pressure, $P_{t,2}$. Here, the phase transition occurs at densities somewhat below the central density of the maximum mass configuration of the purely hadronic star (black dot). For this regime, the value of $P_t$ ensures that the Seidov limit is crossed for almost all explored values of $\Delta\varepsilon$; a minority of cases with low enough $\Delta\varepsilon$ values produce Type~T morphologies, always with short or almost negligible US hybrid branches ($\Delta \varepsilon_1$ case in the figure). Moreover, the transition pressure is high enough that unconditional stability is never recovered beyond the phase transition (corresponding to topology~$A$ or $C$  of Fig.~3 of Ref.~\cite{Alford:2013aca}). Consequently, few US hybrid configurations exist for the parameter combinations shown, and the morphologies are mostly Type~W: the entire post-transition sequence is a monotonically descending SSHS branch.

The branch length increases monotonically with $\Delta\varepsilon$, from a short dark-blue segment at $\Delta\varepsilon_1$ to a long red branch at $\Delta\varepsilon_5$. The role of $c_s^2$ is now reversed compared to the low-$P_t$ regime: a higher speed of sound \emph{lengthens} the SSHS branch. Since unconditional stability can not be recovered regardless of $c_s^2$ in this regime, the only effect of a stiffer quark phase is to sustain the hybrid sequence over a wider range of central densities before the fundamental mode becomes dynamically unstable, thereby delaying the destabilization and extending the branch. The longest SSHS branches in our entire sample arise for Type~W in this high-$P_t$, high-$\Delta\varepsilon$ regime.

\paragraph*{Summary of the dual role of $c_s^2$.} The schematic cases of Fig.~\ref{fig:mraio_selec} reveal that the speed of sound in the quark phase plays a qualitatively different role depending on the morphological type. In the Type~B (and TB) regime at low $P_t$, where the SSHS branch bridges two US segments, a higher $c_s^2$ facilitates the recovery of unconditional stability and thereby shortens the bridge. In the Type~W regime at high $P_t$, where unconditional stability is never recovered, a higher $c_s^2$ instead sustains the hybrid sequence over a broader density range, lengthening the branch. The Type~T regime exhibits no uniform trend: increasing $c_s^2$ shortens the branch at low pressures but lengthens it at high pressures.

\begin{figure*}[t!]
\centering
\includegraphics[width=\linewidth]{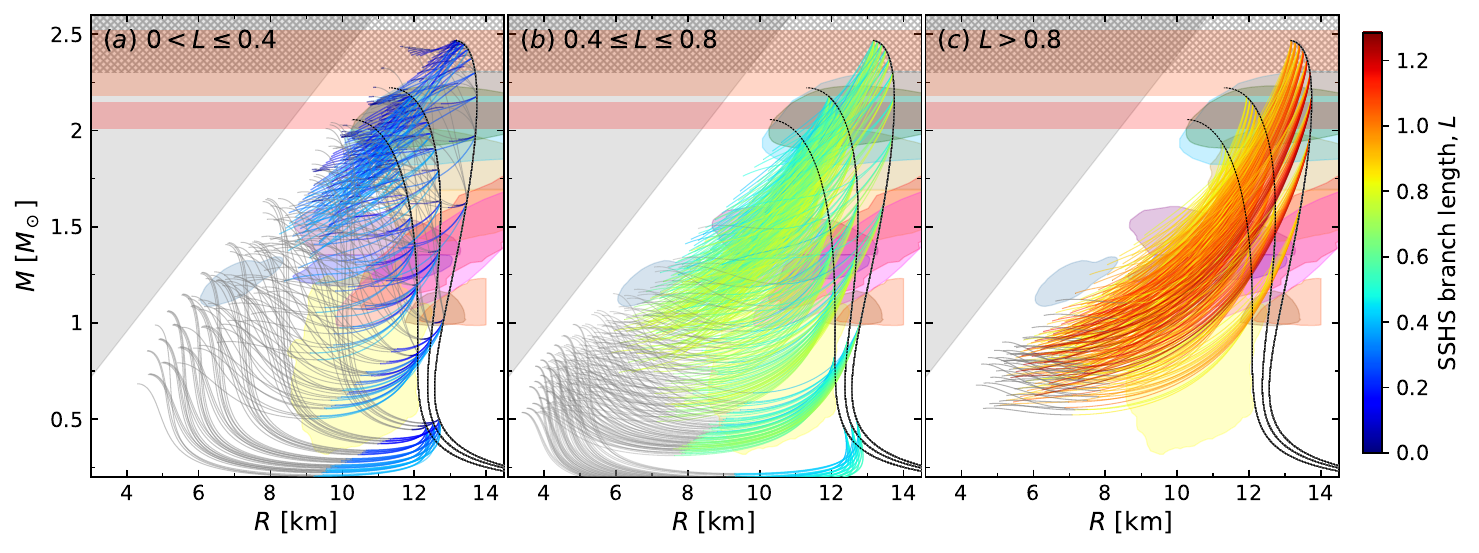}
\caption{Same data as Fig.~\ref{fig:mrlenghts_multi_had}, now grouped by the value of the SSHS branch length: $0 < L \leq 0.4$~(a), $0.4 < L \leq 0.8$~(b), and $L > 0.8$~(c), regardless of the hadronic EoS. This rearrangement reveals how the morphological types distribute across the three $L$ ranges: predominantly Type~B, T, and TB at low $L$~(a); Type~B, TB, and W at intermediate $L$~(b); and mostly Type~W, with a few Type~B configurations, at high $L$~(c).}
\label{fig:mrlenghts_multi_long}
\end{figure*}

\subsection{Impact of the hadronic EoS on the SSHS branch}
\label{sec:impact_had}

With the morphological types identified, and qualitatively illustrated by the schematic cases in Figs.~\ref{fig:sketch} and ~\ref{fig:mraio_selec}, we now examine the full sample of 1521 hybrid EoSs. We first analyze the impact of the hadronic EoS on the SSHS branch properties in this subsection. We then characterize the morphological diversity as a function of the branch length, $L$, in Sec.~\ref{sec:branch_diversity}.

Figure~\ref{fig:mrlenghts_multi_had} presents the $M$-$R$ relationships for all hybrid EoSs, separated by the underlying hadronic model: \soft [panel~(a)], \inter [panel~(b)], and \stiff [panel~(c)]. The purely hadronic sequences are shown as black curves; gray segments correspond to US configurations; and the first SSHS branch is colored according to $L$.

The maximum attainable value of $L$ increases systematically with the stiffness of the hadronic EoS. In the \soft case, the SSHS branches are predominantly short, with $L$ concentrated in the blue end of the color scale (\mbox{$L \lesssim 0.4$-$0.6$}); only a limited number of parameter combinations produce branches reaching $L \sim 0.8$. In the \inter case, branches with $L \sim 0.6$-$0.8$ become common, and the longest branches approach $L \sim 1.0$. The \stiff case yields the longest branches overall, with numerous configurations reaching $L > 1.0$ and the global maximum exceeding $L \simeq 1.2$. This trend is physically natural: a stiffer hadronic phase produces stars with larger radii (up to $\sim 14$~km for the \stiff EoS compared to $\sim 12$~km for the \soft one), providing a wider span in the $M$-$R$ plane over which the SSHS branch can develop once the phase transition softens the EoS and drives the sequence toward smaller radii.

Three morphological types (W, B, and T) are clearly present across the three panels, together with sets that exhibit no SSHS branch. Type~TB configurations are infrequent and not easily distinguishable from Type~B; this potential ambiguity between two closely related morphologies does not affect the main results of our work. The relative prevalence of each morphological type and, crucially, its ability to satisfy the $M_{\max} \geq 2.01\,M_\odot$ constraint vary significantly with the hadronic EoS. Before turning to the full set of observational constraints in Sec.~\ref{sec:results2_astro}, we examine the impact of this single requirement, which is the most robust among current astrophysical bounds and by itself eliminates a large fraction of the constructed models. We discuss each panel in turn.

In the \soft case [Fig.~\ref{fig:mrlenghts_multi_had}~(a)], the purely hadronic EoS only marginally exceeds $2\,M_\odot$, leaving very little room for the softening introduced by the phase transition. As a result, the majority of hybrid configurations fail the maximum mass constraint. This is particularly severe for Type~W branches (warm colors) because most of them depart from a maximum mass that already falls below $2\,M_\odot$. Type~B configurations (blue tones) are also numerous, but even in these cases, the maximum mass of the US hybrid branch ---softened by the presence of the quark core--- often fails to reach the $2\,M_\odot$ threshold. Overall, the \soft hadronic EoS produces very few astrophysically viable hybrid configurations, regardless of the morphological type.

In the \inter case [Fig.~\ref{fig:mrlenghts_multi_had}~(b)], the purely hadronic EoS reaches $M_{\max} \sim 2.2$-$2.3\,M_\odot$, providing a larger margin. Some Type~T configurations, with short SSHS tails appearing beyond an US hybrid branch, have their maximum mass well above $2\,M_\odot$. Type~W branches also become viable in many cases, provided that their departure point ---the maximum mass of the sequence--- remains above $2\,M_\odot$. The most constrained morphology is Type~B, where the SSHS segment acts as an intermediate bridge followed by an US hybrid branch: the maximum mass of this hybrid branch depends sensitively on the balance between $\Delta\varepsilon$ (which suppresses it) and $c_s^2$ (which supports it), and only marginal Type~B configurations succeed in reaching $2\,M_\odot$.

In the \stiff case [Fig.~\ref{fig:mrlenghts_multi_had}~(c)], the pure hadronic EoS reaches $M_{\max} \sim 2.4\,M_\odot$, and all morphological types can exceed $2\,M_\odot$, though with varying degrees of ease. The morphology that most readily satisfies the maximum mass constraint is Type~W, in which the SSHS branch originates directly at the hadronic turning point with no subsequent recovery of unconditional stability. A smaller number of configurations correspond to Type~T, with short SSHS tails appearing beyond an US hybrid branch that extends up to the maximum mass. Very few Type~B, which is the most penalized by the density jump, retain $M_{\max} \geq 2\,M_\odot$ mainly due to the large baseline provided by the stiff hadronic phase. Some configurations with moderate $\Delta\varepsilon$, however, pass the threshold only marginally. The astrophysical viability of the \stiff case is further constrained by additional observational bounds, which will be discussed in detail in Sec.~\ref{sec:results2_astro}.

A general pattern emerges from this comparison. Among the different morphological types, Type~B (and TB) ---in which the SSHS segment bridges the hadronic branch to an US hybrid branch whose maximum mass determines $M_{\max}$--- is systematically the most constrained by the $2\,M_\odot$ requirement. The maximum mass of the post-bridge hybrid branch is inevitably reduced by the softening from the phase transition, and this reduction is only partially compensated by the stiffness of the quark phase. By contrast, Type~W branches, whose maximum mass is set by the hadronic portion of the sequence, are less sensitive to the density jump and become viable whenever the hadronic EoS provides a sufficiently high baseline $M_{\max}$.

\subsection{Morphological trends with branch length}
\label{sec:branch_diversity}

A complementary perspective on the full sample is provided by Fig.~\ref{fig:mrlenghts_multi_long}, where all hybrid EoSs are regrouped by the value of $L$, regardless of the hadronic model. For clarity, the $L=0$ cases are omitted from this figure.

Figure~\ref{fig:mrlenghts_multi_long}~(a) ($0 < L \leq 0.4$) exhibits the richest morphological diversity. The dominant population consists of Type~B (and a few TB) configurations with short SSHS bridges (blue tones) connecting the stable hadronic branch to an US hybrid branch at lower masses. The extent of the subsequent US hybrid branch varies considerably: in some cases, it reaches very compact configurations ($R \sim 4$-$6$~km, $M \sim 0.3$-$0.5\,M_\odot$), while in others it terminates at intermediate masses. Type~T configurations are also present, recognizable as very short SSHS tails (dark blue) beyond the maximum mass of an US hybrid branch. The coexistence of these morphologies within the same range of $L$ reflects the sensitivity of the short-branch regime to the specific combination of $P_t$, $\Delta\varepsilon$, and $c_s^2$.

\begin{figure*}[tb]
\centering
\includegraphics[width=\linewidth]{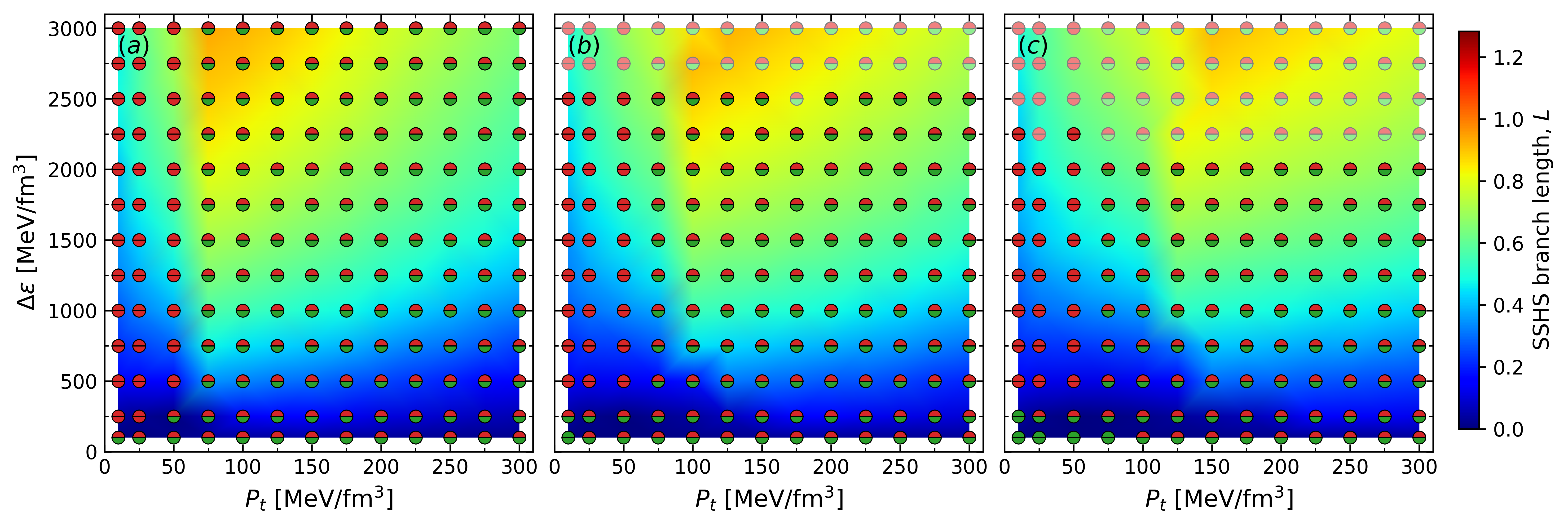}
\caption{Color map of the SSHS branch length $L$ in the $P_t$--$\Delta\varepsilon$ plane for the hadronic \soft EoS, with $c_s^2 = 0.33$~(a), $0.50$~(b), and $0.70$~(c). Bicolored circular markers indicate the constructed parameter sets: the upper (lower) semicircle is green if the $M_{\max} \geq 2.01\,M_\odot$ (GW170817) constraint is satisfied, and red otherwise. Vivid colors denote sets consistent with the pQCD constraint; faded colors denote sets that violate it.}
\label{fig:panel_soft}
\end{figure*}

\begin{figure*}[tb]
\centering
\includegraphics[width=\linewidth]{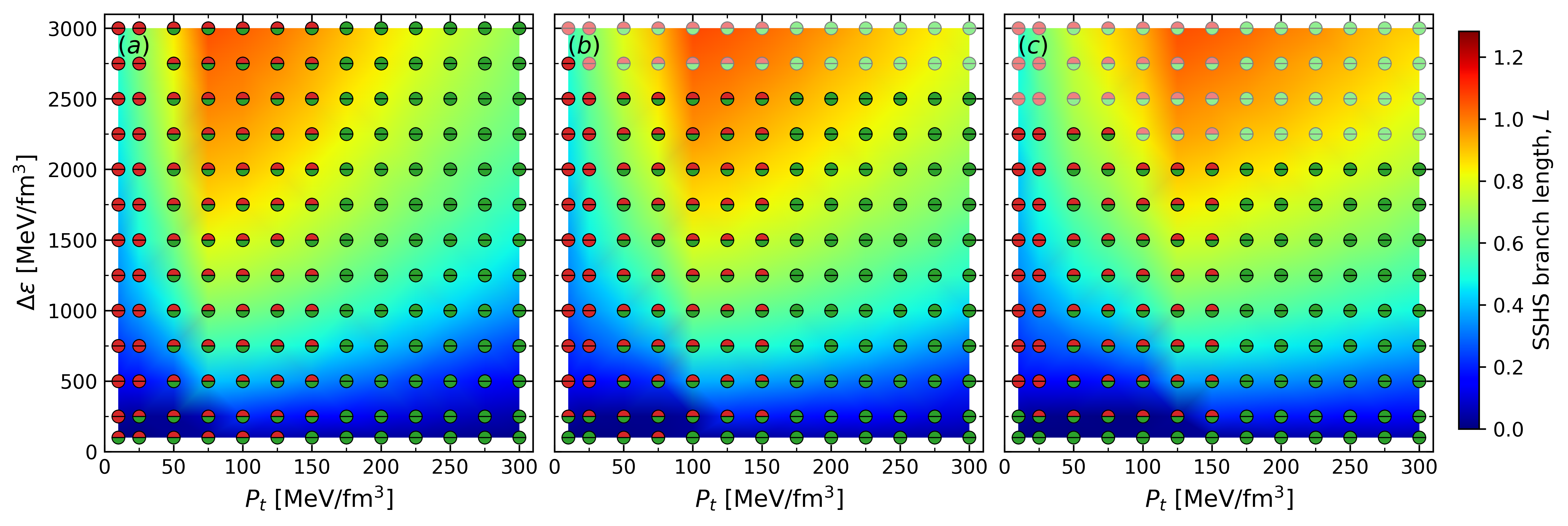}
\caption{Same as Fig.~\ref{fig:panel_soft} but for the hadronic \inter EoS.}
\label{fig:panel_inter}
\end{figure*}

Figure~\ref{fig:mrlenghts_multi_long}~(b) ($0.4 < L \leq 0.8$) displays a transitional regime in which Type~B (or TB) and Type~W morphologies coexist. Type~B configurations with moderate-length SSHS bridges (green and cyan tones) are still present, followed by (hybrid) second US branches of decreasing extent. Alongside them, Type~W branches begin to appear ---SSHS segments of comparable length that descend monotonically with no recovery of unconditional stability. The gradual shift from Type~B to Type~W reflects the progressive strengthening of the phase transition with increasing $\Delta\varepsilon$: the post-bridge US hybrid branch becomes shorter and eventually vanishes, leaving only the monotonically descending SSHS sequence. In this regime, the SSHS branch sweeps through the central region of the $M$-$R$ plane ($R \sim 8$-$12$~km, $M \sim 0.8$-$1.5\,M_\odot$).

Figure~\ref{fig:mrlenghts_multi_long}~(c) ($L > 0.8$) is dominated by Type~W morphologies. The SSHS branches ---colored in yellow, orange, and red--- extend with a pronounced negative slope ($\partial M / \partial R < 0$), producing a distinctive ``waterfall'' appearance; their endpoint is set by the dynamical destabilization of the fundamental oscillation mode. Some curves in this panel, associated with lower transition pressures, do exhibit a recovery of unconditional stability at low masses (Type~B with very long bridges); however, these configurations fail the $M_{\max} \geq 2.01\,M_\odot$ constraint and are therefore not astrophysically viable ---a point we return to in Sec.~\ref{sec:results2_astro}. The astrophysically relevant long branches are Type~W configurations that connect the high-mass regime with the low-mass, compact region of the $M$-$R$ plane through a single continuous stellar sequence, potentially accommodating both massive pulsars and anomalously light compact objects within the same family of stable configurations.

\section{Astrophysical constraints}
\label{sec:results2_astro}

\begin{figure*}[tb]
\centering
\includegraphics[width=\linewidth]{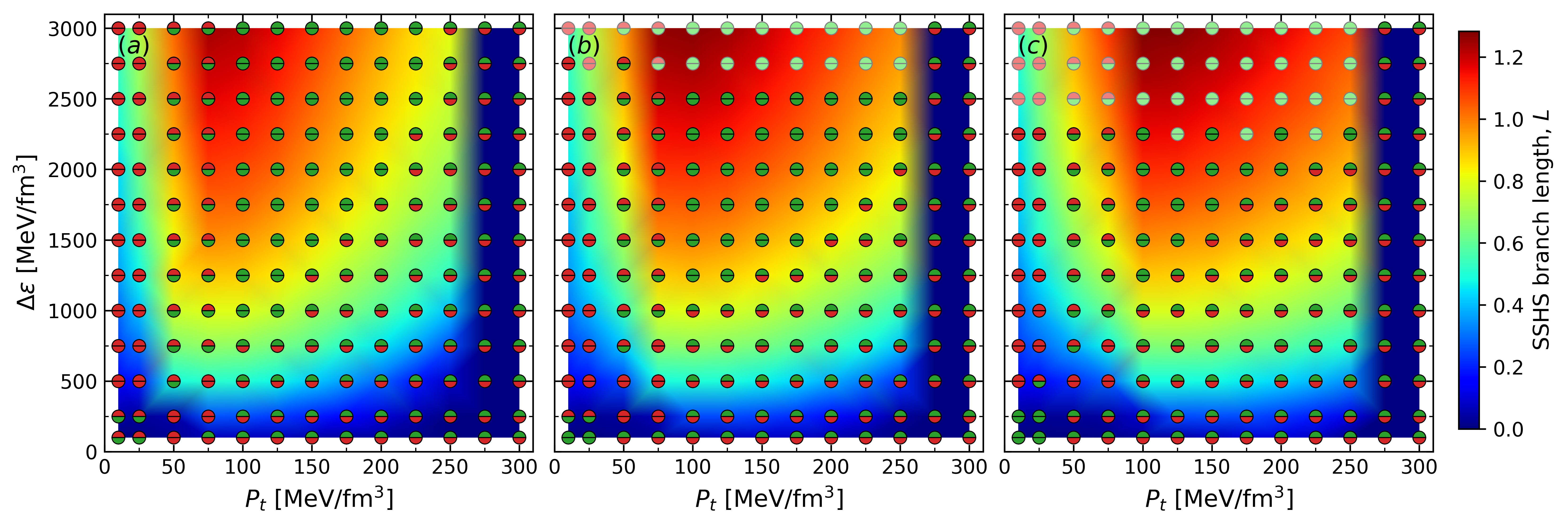}
\caption{Same as Fig.~\ref{fig:panel_soft} but for the hadronic \stiff EoS.}
\label{fig:panel_stiff}
\end{figure*}

In this section, we confront the full sample of hybrid EoSs with current astrophysical constraints. We filter our sets using two well-established requirements: the maximum mass bound $M_{\max} \geq 2.01\,M_\odot$, and consistency with the masses and radii inferred from GW170817. Several considerations motivate this choice. First, there is a significant overlap among the regions excluded by different constraints in the $M$-$R$ plane, so that satisfying this pair effectively ensures compatibility with the remaining ones, including NICER mass-radius estimates. Second, more stringent constraints ---such as the HESS~J1731$-$347~\cite{lugones:2023ama} and XTE~J1814$-$338~\cite{Mariani:2026hns} observations--- have been addressed in our previous work and would introduce additional complexity without substantial gain, given the filtering already applied.
Third, the upper bound $M_{\max} \lesssim 2.3\,M_\odot$ inferred by \citet{Shibata:2019ctb} from GW170817 and kilonova modeling is satisfied for the vast majority of our viable sets. In the \stiff hadronic case, however, a non-negligible subset of configurations with transition pressures $P_t \simeq 150$--$250~\mathrm{MeV/fm^3}$ marginally exceeds this bound. Given that it relies on specific assumptions about merger dynamics and the modeling of kilonova ejecta, we treat it as a model-dependent constraint and do not impose it as a strict selection criterion in this work. We additionally track consistency with the high-density pQCD constraint~\cite{Annala:2020efq}, as described below.

Figures~\ref{fig:panel_soft}, \ref{fig:panel_inter}, and \ref{fig:panel_stiff} present color-maps of $L$ in the \mbox{$P_t$-$\Delta\varepsilon$} plane for the \soft, \inter, and \stiff hadronic EoSs, respectively, with three panels per figure corresponding to $c_s^2 = 0.33$ (a), $0.5$ (b), and $0.7$ (c). These nine panels display the dependence of $L$ on all model parameters in a non-degenerate and exhaustive way. Each panel also shows the EoS sample through bicolored circular markers: the upper (lower) semicircle is green if the $2.01\,M_\odot$ (GW170817) constraint is satisfied and red otherwise. Sets displayed with vivid colors satisfy the pQCD constraint; those with faded colors do not. Fully-green vivid markers, therefore, represent the subset of our sample that meets all current microphysical and astrophysical constraints simultaneously.

\subsection{Dependence of $L$ on the EoS parameters}

A comparison across the three figures confirms that stiffer hadronic EoSs produce significantly longer SSHS branches. For a given $c_s^2$, neither the location of the maximum of $L$ nor the overall color-gradient structure across the $P_t$-$\Delta\varepsilon$ plane changes appreciably when different hadronic EoSs are used. The only exception is the \stiff EoS at $P_t \gtrsim 255~\mathrm{MeV/fm^3}$, where $P_t$ exceeds $P_c^{M_{\max}}$ and $L = 0$ is assigned by construction (see Sec.~\ref{sec:quarks}).

Variations in $c_s^2$ modify $L$ in the manner anticipated by the schematic analysis of Fig.~\ref{fig:mraio_selec}: they do not alter the color-map structure qualitatively, but shift the location of the maximum of $L$ toward higher pressures. The maximum value of $L$ attained also depends on the hadronic EoS: it is found in panel~(a) for the \soft case, panel~(b) for \inter, and panel~(c) for \stiff. The global maximum across the entire sample is $L = 1.28$, corresponding to a Type~W configuration with the \stiff hadronic EoS and $c_s^2 = 0.7$; this set satisfies the astrophysical constraints but not the pQCD bound.

\subsection{Constraint filtering}

The outcome of the constraint filtering depends strongly on the hadronic EoS. The \soft case lacks a sufficiently high $M_{\max}$, making it difficult to retain the $2.01\,M_\odot$ bound once the softening from the phase transition is included. Only a limited number of sets with low $P_t$, low $\Delta\varepsilon$, and high $c_s^2$ remain viable, all corresponding to below-Seidov-limit morphologies (Type~T). The \inter and \stiff EoSs do not suffer from this limitation, as many of their sets exceed $2.01\,M_\odot$. The two cases differ, however, in the way the GW170817 bound is met: the \inter EoS satisfies it by construction through its hadronic branch, so that \inter sets can fulfill it via hadronic, USHS, or SSHS configurations (Types~W, T, and marginally B or TB). The \stiff EoS, by contrast, can meet this bound only through hybrid branches, the vast majority of which are Type~W.

This difference is reflected in the distribution of fully-green markers across the figures. The \inter case displays a broad, roughly rectangular region of viable sets at high pressures across all three panels. The \stiff case exhibits a less numerous, approximately triangular distribution concentrated at intermediate pressures and large $\Delta\varepsilon$. In both cases, the $c_s^2 = 0.5$ and $0.7$ panels show a progressive reduction in the number of viable sets, as higher $c_s^2$ values lead to increasingly frequent violations of the pQCD bound at large $\Delta\varepsilon$.

\subsection{Central densities and the pQCD regime}

Among the pQCD-consistent filtered sets, we find a correlation between $L$ and the central baryon number density of the terminal (last stable) configuration, $n_{B,c}^{\rm{term}}$. Notably, the SSHS configurations in these sets can reach central densities of up to $\sim 60\,n_0$, far exceeding the values typically attained in purely hadronic stars (see Table~\ref{tabla:eos_had}) and extending toward the regime where pQCD calculations become relevant. If the dynamical stability of HSs is indeed regulated by slow phase conversion, this class of objects could provide a potential window into density ranges in compact star cores that  are not usually considered accessible.

\subsection{The $n_{B,t}$-$\Delta n_B$ plane}

\begin{figure*}[tbh]
\centering
\includegraphics[width=0.95\linewidth]{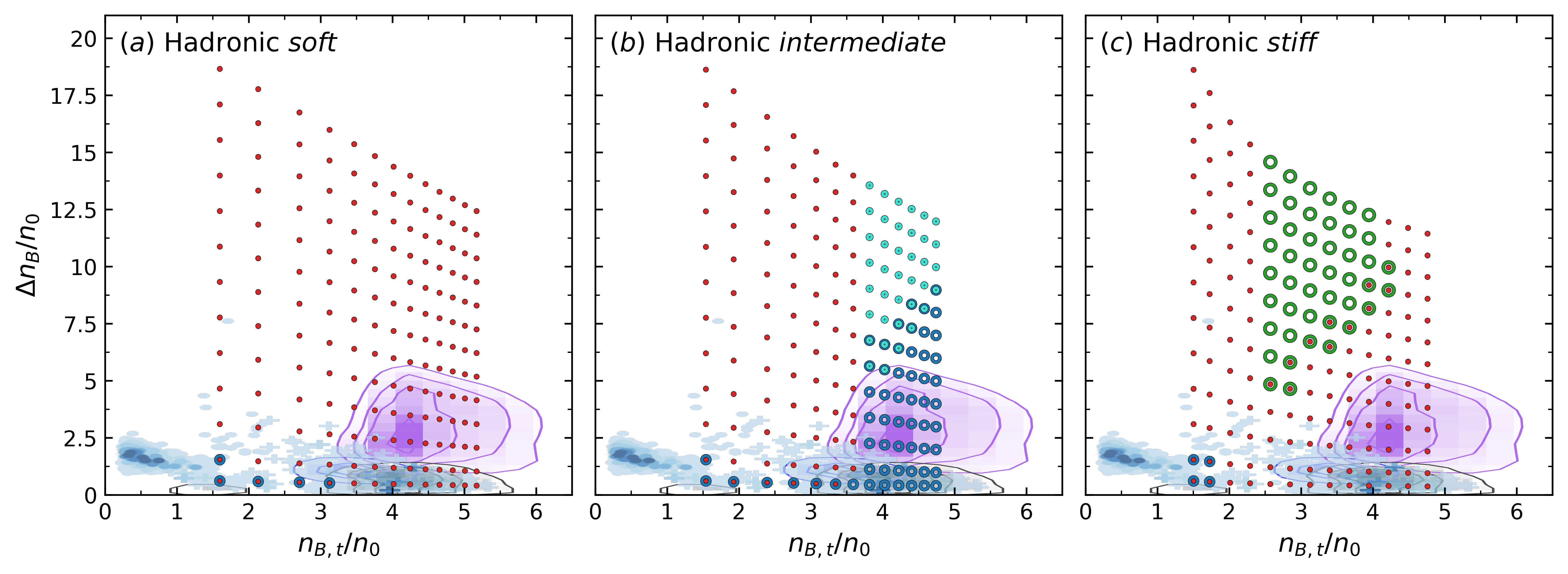}
\caption{Hadronic baryon number density at the phase transition, $n_{B,t}$, versus baryon number density jump, $\Delta n_B$ (both in units of $n_0$), for all constructed sets: \soft~(a), \inter~(b), and \stiff~(c) hadronic EoSs. See text for details on the marker color coding. Confidence-level contours and the background distribution of blue points correspond to results from \citet{Komoltsev:2024fop} and \citet{Blomqvist:2025sma}, respectively.}
\label{fig:panel_nb}
\end{figure*}

Figure~\ref{fig:panel_nb} presents the EoS sampling in the plane of hadronic baryon number density at the phase transition, $n_{B,t}$, versus the baryon number density jump, $\Delta n_B$, both in units of $n_0 = 0.16$~fm$^{-3}$. Panels~(a), (b), and (c) correspond to the \soft, \inter, and \stiff hadronic EoSs, respectively. Since $c_s^2$ does not enter the quantities plotted, some degeneracy in the set positions arises within each panel; to make all sets visible, we use concentric colored markers. Red markers denote sets that fail at least one of the three constraints (pQCD, $2.01\,M_\odot$, or GW170817). Blue, cyan, and green rings denote sets satisfying all three, distinguished by how the GW170817 bound is met: exclusively through US branches (blue), through both US and SSHS branches (cyan), or exclusively through the SSHS branch (green). For comparison, we overlay the confidence-level contours of \citet{Komoltsev:2024fop} and the posterior distribution of \citet{Blomqvist:2025sma}.

The \soft case [Fig.~\ref{fig:panel_nb}~(a)] yields only marginally viable sets at very low transition densities, meeting the GW170817 bound through US hybrid branches below the Seidov limit. The \inter case [Fig.~\ref{fig:panel_nb}~(b)] meets this bound by construction through its hadronic branch, and accordingly displays blue and cyan markers but no green ones; cyan sets ---in which an SSHS branch also traverses the GW170817 region--- appear at larger $\Delta n_B$, corresponding to longer Type~W branches. The \stiff case [Fig.~\ref{fig:panel_nb}~(c)] is the only one exhibiting green markers: these sets exploit the long Type~W SSHS branches to meet the GW170817 bound despite the \textit{a priori} disfavored hadronic branch.

Compared with the analyses of \citet{Komoltsev:2024fop} and \citet{Blomqvist:2025sma}, which adopt the rapid conversion framework, our results show good overall agreement when restricted to the same scenario, beyond the natural differences arising from the diversity of density values explored in each work. When SSHSs are included, however, an entirely new region of the $n_{B,t}$--$\Delta n_B$ plane (green markers) emerges as viable. This finding indicates that stiff hadronic EoSs should not be discarded prematurely ---in contrast to the conclusions of the aforementioned studies, which strongly disfavor such models.

\section{Summary and discussion}
\label{sec:conclus}

In this work, we have introduced and systematically studied the length $L$ of the SSHS branch as a quantitative measure of the extended stability region that arises when the hadron-quark phase conversion is slow compared to the radial oscillation timescale. By combining generalized piecewise-polytropic hadronic EoSs with constant-speed-of-sound quark matter, and by scanning a broad set of transition pressures, energy-density jumps, and quark-matter stiffness, we obtained a systematic mapping of the HS parameter space under the slow conversion hypothesis. This analysis shows that slow conversion does not simply enlarge the stable HS region quantitatively, but qualitatively modifies the phenomenology of HS sequences in the $M$-$R$ plane.

A central result of this work is the identification of four morphological types for the SSHS branch ---waterfall (W), bridge (B), tail (T), and tail-bridge (TB)--- which are specific to the slow-conversion framework. In contrast to the rapid-conversion case, where the topology of hybrid branches is controlled by the Seidov criterion and the classification introduced in Ref.~\cite{Alford:2013aca}, slow conversion stabilizes configurations that would otherwise be radially unstable and gives rise to a distinct morphological structure. Given a particular hadronic EoS, the morphology is governed primarily by the transition pressure $P_t$ and the magnitude of the energy-density jump $\Delta\varepsilon$. Low transition pressures and large jumps favor Type~B configurations, while high transition pressures favor Type~W branches; Type~T occurs for sufficiently small $\Delta\varepsilon$, whereas Type~TB appears only in a comparatively narrow region of parameter space.

The branch length $L$ provides a useful way of organizing this morphological diversity. Short SSHS branches ($L \lesssim 0.4$) display the largest variety of morphologies, with Types~B, TB, and T coexisting within the same range of~$L$. At intermediate lengths ($0.4 \lesssim L \lesssim 0.8$), a transitional regime emerges in which Type~B and Type~W configurations coexist, reflecting the progressive strengthening of the phase transition as $\Delta\varepsilon$ increases. Long branches ($L \gtrsim 0.8$) are overwhelmingly dominated by Type~W configurations. This hierarchy becomes particularly relevant once observational constraints are imposed. Among the different morphologies, Type~B and TB are the most strongly penalized by the $M_{\max} \geq 2.01\,M_\odot$ constraint because the post-transition hybrid branch must recover a sufficiently large maximum mass after the softening induced by the phase transition. As a result, the long SSHS branches that remain astrophysically viable are predominantly of Type~W. These are also the configurations most capable of connecting, within a single continuous stable sequence, the high-mass sector of the $M$-$R$ plane to a low-mass compact region.

Another robust result is the dual role of the quark-matter speed of sound. In the Type~B regime, increasing $c_s^2$ shortens the SSHS bridge by facilitating the recovery of unconditional stability. In the Type~W regime, by contrast, a larger $c_s^2$ extends the slow-stable branch by sustaining the hybrid sequence over a broader density interval before destabilization sets in. The maximum attainable length $L$ also increases with the stiffness of the hadronic EoS, showing that stiff hadronic backgrounds are the most favorable for producing extended SSHS branches; the global maximum found in our sample is $L = 1.28$, corresponding to a Type~W configuration with the \stiff hadronic EoS and $c_s^2 = 0.7$. In this case, the SSHS branch spans a mass range of over $1.5\,M_\odot$ and a radius range of about $6$~km. At the same time, the longest branches do not necessarily coincide with the most conservative microphysical parameter sets: this global maximum satisfies the astrophysical constraints considered here but not the pQCD bound, illustrating that the region of maximum SSHS extent and the region of full microphysical viability do not, in general, coincide.

Filtering the sample through the $M_{\max} \geq 2.01\,M_\odot$ and GW170817 constraints reveals a strong dependence on the hadronic EoS. The \soft model leaves little room for the softening associated with the phase transition and therefore yields only a small number of viable hybrid configurations, all of Type~T. The \inter model retains a broad viable region, particularly at high transition pressures; because its hadronic branch already satisfies the GW170817 bound, viable sets can fulfill this constraint through hadronic, USHS, or SSHS configurations. The \stiff model presents the most distinctive behavior: its purely hadronic branch does not satisfy the GW170817 bound, so that compliance with this constraint requires the softening provided by the hybrid branches. For the vast majority of viable \stiff sets, the GW170817 region is reached exclusively through the SSHS branch (Type~W), making slow-conversion stability essential for the viability of stiff hadronic EoSs. In this sense, our results reinforce and extend the conclusions of Ref.~\cite{lugones:2023ama}: the viability of stiff hadronic EoSs depends crucially on the dynamical nature of the hadron-quark conversion process.

The comparison with the analyses of \citet{Komoltsev:2024fop} and \citet{Blomqvist:2025sma} in the $n_{B,t}$--$\Delta n_B$ plane shows good overall consistency when only rapid stability is considered. Once SSHSs are included, however, a new viable region appears ---~occupied exclusively by \stiff-EoS sets whose GW170817 compliance relies on the SSHS branch--- that is entirely absent under the rapid-conversion hypothesis. This finding has a noteworthy implication for the density regime probed by compact stars: the pQCD-consistent SSHS configurations within this new region can reach central densities up to $\sim 60\,n_0$, suggesting that the slow conversion hypothesis could open an observational window into density regimes not usually considered accessible.

A few aspects of our analysis deserve further comment. First, throughout this paper, the SSHS branch refers exclusively to the first slow-stable segment encountered along a sequence of increasing central density. Secondary slow-stable segments may appear beyond the maximum mass of a subsequent US hybrid branch in Type~B and TB morphologies; such secondary segments are excluded from the present study. Likewise, scenarios in which the phase transition occurs beyond the hadronic maximum-mass configuration ---potentially giving rise to disconnected slow hybrid branches--- are not considered here and require a dedicated analysis. Second, the CSS parametrization assumes a constant speed of sound in the quark phase, whereas microscopic models generically predict a density-dependent $c_s^2$. The use of CSS is motivated by its widespread adoption as a flexible and minimal parametrization that captures the essential features of a first-order phase transition: the location of the transition pressure, the magnitude of the energy-density jump, and the average stiffness of the post-transition matter. Within this framework, our morphological classification (W, B, T, TB) is governed by the local behavior of the EoS at and immediately beyond the phase transition \mbox{---primarily} $P_t$ and $\Delta\varepsilon$, with $c_s^2$ playing a secondary role---, and is therefore expected to remain meaningful for non-CSS EoSs that share these qualitative features. The dual role of $c_s^2$ in shortening Type~B branches and lengthening Type~W branches, however, has been established here only for constant $c_s^2$; for density-dependent quark EoSs, an analogous role would presumably be played by some effective average stiffness, but the precise mapping is not straightforward and deserves a dedicated study. Likewise, the quantitative results ---the specific values of $L$, the location of the global maximum, and the boundaries of the viable regions in the $P_t$-$\Delta\varepsilon$ plane--- are tied to the CSS assumption and may shift under alternative parametrizations. A preliminary indication that Type~W branches persist for non-CSS quark EoSs can be found in Ref.~\cite{Mariani:2026hns}, but a systematic comparison across multiple microscopic quark models is left for a future work.

Forthcoming observational milestones may make it possible to distinguish SSHSs from US stars. In addition to the SSHS hypothesis, several alternative scenarios have been proposed to account for the growing body of NS observations, including conventional hadronic stars, strange quark stars, and the two-families scenario~\cite{Drago:2016tso1}. A first diagnostic relies on the determination of masses and radii. As discussed in Ref.~\cite{Mariani:2026hns}, the mass and radius of extremely compact objects such as XTE~J1814$-$338 remain controversial; were these estimates robustly confirmed, they would provide reliable evidence for the existence of HSs (or strange quark stars), although they could not by themselves discriminate among conversion-rate scenarios. Even for less extreme objects, mass and radius measurements more precise than those currently available could help disentangle the possible conversion scenarios, identify the favored stable branches, and thereby place additional constraints on the underlying theoretical models. A second, complementary diagnostic relies on non-radial oscillation modes. Because $g$-modes can be excited only in the slow-conversion regime, their detection would provide a strong hint of both a sharp hadron-quark phase transition and a slow conversion rate between the two phases (see, e.g., Refs.~\cite{tonetto:2020dgm,Mariani:2022omh}). Likewise, departures from the universal relations obeyed by non-radial oscillation modes may be interpreted as signatures of SSHSs (see, e.g., Refs.~\cite{Ranea:2022bou,Ranea:2023auq}). Neither asteroseismic strategy is straightforward to implement, however, since the unambiguous identification of an individual mode is hindered by the strong degeneracy among mode frequencies.

Taken together, our results show that the slow-conversion hypothesis qualitatively reshapes the structure of HS sequences, the morphology of the stable branches, and the range of hadronic EoSs that remain astrophysically viable. If future observations were to favor the existence of extended stable  hybrid branches, our results would provide guidance ---within the CSS framework--- on the phase-transition parameters most compatible with such a scenario. As discussed in Sec.~\ref{sec:results2_astro}, long SSHS branches that simultaneously satisfy the $M_{\max} \geq 2.01\,M_\odot$ and GW170817 constraints are predominantly of Type~W, since Type~B configurations with comparable $L$ are systematically penalized by the maximum-mass requirement. Because Type~W morphologies arise at sufficiently high transition pressures ---where unconditional stability is not recovered beyond the phase transition--- our analysis points toward intermediate-to-high values of $P_t$ combined with comparatively large energy-density jumps, with the average stiffness of the quark-matter sector also playing a relevant role in determining both the branch morphology and its extent.

\section*{Acknowledgments}
MM, MGO and IFR-S acknowledge UNLP and CONICET (Argentina) for financial support under grants G187, G009 and PIP 0169. GL acknowledges the partial financial support from the Brazilian agency CNPq (grant 316844/2021-7).

\bibliography{references}

@ARTICLE{Zhang:2026cas-arXiv,
       author = {{Zhang}, Chen},
        title = "{Can a Slow and Strong Phase Transition in Neutron Stars Relieve Major Compact-Star Observation Tensions?}",
      journal = {arXiv e-prints},
         year = 2026,
        month = jun,
          eid = {arXiv:2606.17231},
        pages = {arXiv:2606.17231},
          doi = {10.48550/arXiv.2606.17231},
archivePrefix = {arXiv},
       eprint = {2606.17231},
 primaryClass = {astro-ph.HE},
       adsurl = {https://ui.adsabs.harvard.edu/abs/2026arXiv260617231Z}
}

@article{Gholami:2025aco,
       author = {{Gholami}, Hosein and {Rather}, Ishfaq Ahmad and {Hofmann}, Marco and {Buballa}, Michael and {Schaffner-Bielich}, J{\"u}rgen},
        title = "{Astrophysical constraints on color-superconducting phases in compact stars within the RG-consistent NJL model}",
      journal = {\prd},
         year = 2025,
        month = may,
       volume = {111},
       number = {10},
          eid = {103034},
        pages = {103034},
          doi = {10.1103/PhysRevD.111.103034},
archivePrefix = {arXiv},
       eprint = {2411.04064},
 primaryClass = {hep-ph},
       adsurl = {https://ui.adsabs.harvard.edu/abs/2025PhRvD.111j3034G}
}

@article{Mclerran:2019qma,
       author = {{McLerran}, Larry and {Reddy}, Sanjay},
        title = "{Quarkyonic Matter and Neutron Stars}",
      journal = {\prl},
         year = 2019,
        month = mar,
       volume = {122},
       number = {12},
          eid = {122701},
        pages = {122701},
          doi = {10.1103/PhysRevLett.122.122701},
archivePrefix = {arXiv},
       eprint = {1811.12503},
 primaryClass = {nucl-th},
       adsurl = {https://ui.adsabs.harvard.edu/abs/2019PhRvL.122l2701M}
}

@book{harrison1965gravitation,
  title={Gravitation Theory and Gravitational Collapse},
  author={Harrison, B.K. and Thorne, K.S. and Wakano, M. and Wheeler, J.A.},
  isbn={9780226318028},
  lccn={65017293},
  url={https://books.google.com.br/books?id=0V7AOwAACAAJ},
  year={1965},
  publisher={University of Chicago Press}
}

@article{mariani:2019mhs,
       author = {{Mariani}, Mauro and {Orsaria}, Milva G. and {Ranea-Sandoval}, Ignacio F. and {Lugones}, Germ{\'a}n},
        title = "{Magnetized hybrid stars: effects of slow and rapid phase transitions at the quark-hadron interface}",
      journal = {\mnras},
         year = 2019,
        month = nov,
       volume = {489},
       number = {3},
        pages = {4261-4277},
          doi = {10.1093/mnras/stz2392},
archivePrefix = {arXiv},
       eprint = {1909.08661},
 primaryClass = {astro-ph.HE},
       adsurl = {https://ui.adsabs.harvard.edu/abs/2019MNRAS.489.4261M}
}

@ARTICLE{Ranea:2022bou,
       author = {{Ranea-Sandoval}, Ignacio F. and {Guilera}, Octavio M. and {Mariani}, Mauro and {Lugones}, Germ{\'a}n},
        title = "{Breaking of universal relationships of axial w I modes in hybrid stars: Rapid and slow hadron-quark conversion scenarios}",
      journal = {\prd},
         year = 2022,
        month = aug,
       volume = {106},
       number = {4},
          eid = {043025},
        pages = {043025},
          doi = {10.1103/PhysRevD.106.043025},
archivePrefix = {arXiv},
       eprint = {2208.07667},
 primaryClass = {astro-ph.HE},
       adsurl = {https://ui.adsabs.harvard.edu/abs/2022PhRvD.106d3025R}
}

@ARTICLE{Ranea:2023cmr,
       author = {{Ranea-Sandoval}, Ignacio F. and {Mariani}, Mauro and {Lugones}, Germ{\'a}n and {Guilera}, Octavio M.},
        title = "{Constraining mass, radius, and tidal deformability of compact stars with axial wI modes: new universal relations including slow stable hybrid stars}",
      journal = {\mnras},
         year = 2023,
        month = feb,
       volume = {519},
       number = {2},
        pages = {3194-3200},
          doi = {10.1093/mnras/stac3780},
archivePrefix = {arXiv},
       eprint = {2212.10514},
 primaryClass = {astro-ph.HE},
       adsurl = {https://ui.adsabs.harvard.edu/abs/2023MNRAS.519.3194R}
}

@ARTICLE{Ranea:2023auq,
       author = {{Ranea-Sandoval}, Ignacio F. and {Mariani}, Mauro and {Celi}, Marcos O. and {Rodr{\'\i}guez}, M. Camila and {Tonetto}, Lucas},
        title = "{Asteroseismology using quadrupolar f -modes revisited: Breaking of universal relationships in the slow hadron-quark conversion scenario}",
      journal = {\prd},
         year = 2023,
        month = jun,
       volume = {107},
       number = {12},
          eid = {123028},
        pages = {123028},
          doi = {10.1103/PhysRevD.107.123028},
archivePrefix = {arXiv},
       eprint = {2306.02823},
 primaryClass = {astro-ph.HE},
       adsurl = {https://ui.adsabs.harvard.edu/abs/2023PhRvD.107l3028R}
}

@ARTICLE{Rau:2023tfo,
       author = {{Rau}, Peter B. and {Sedrakian}, Armen},
        title = "{Two first-order phase transitions in hybrid compact stars: Higher-order multiplet stars, reaction modes, and intermediate conversion speeds}",
      journal = {\prd},
         year = 2023,
        month = may,
       volume = {107},
       number = {10},
          eid = {103042},
        pages = {103042},
          doi = {10.1103/PhysRevD.107.103042},
archivePrefix = {arXiv},
       eprint = {2212.09828},
 primaryClass = {astro-ph.HE},
       adsurl = {https://ui.adsabs.harvard.edu/abs/2023PhRvD.107j3042R}
}

@article{lugones:2023ama,
doi = {10.1088/1475-7516/2023/03/028},
url = {https://doi.org/10.1088/1475-7516/2023/03/028},
year = {2023},
month = {mar},
publisher = {IOP Publishing},
volume = {2023},
number = {03},
pages = {028},
author = {Lugones, Germán and Mariani, Mauro and Ranea-Sandoval, Ignacio F.},
title = {A model-agnostic analysis of hybrid stars with reactive interfaces},
journal = {Journal of Cosmology and Astroparticle Physics},
}

@ARTICLE{Gosh:2024ero,
       author = {{Ghosh}, Sayantan and {Ranjan Mohanty}, Sailesh and {Zhao}, Tianqi and {Kumar}, Bharat},
        title = "{Exploring Radial Oscillations in Slow Stable and Hybrid Neutron Stars}",
      journal = {arXiv e-prints},
         year = 2024,
        month = jan,
          eid = {arXiv:2401.08347},
        pages = {arXiv:2401.08347},
          doi = {10.48550/arXiv.2401.08347},
archivePrefix = {arXiv},
       eprint = {2401.08347},
 primaryClass = {nucl-th},
       adsurl = {https://ui.adsabs.harvard.edu/abs/2024arXiv240108347G}
}

@ARTICLE{Rather:2024roo,
       author = {{Rather}, Ishfaq A. and {Marquez}, Kauan D. and {Backes}, Bet{\^a}nia C. and {Panotopoulos}, Grigoris and {Lopes}, Il{\'\i}dio},
        title = "{Radial oscillations of hybrid stars and neutron stars including delta baryons: the effect of a slow quark phase transition}",
      journal = {\jcap},
         year = 2024,
        month = may,
       volume = {2024},
       number = {5},
          eid = {130},
        pages = {130},
          doi = {10.1088/1475-7516/2024/05/130},
archivePrefix = {arXiv},
       eprint = {2401.07789},
 primaryClass = {nucl-th},
       adsurl = {https://ui.adsabs.harvard.edu/abs/2024JCAP...05..130R}
}

@article{Laskos:2024hsi,
       author = {{Laskos-Patkos}, P. and {Koliogiannis}, P.~S. and {Moustakidis}, Ch. C.},
        title = "{Hybrid stars in light of the HESS J1731-347 remnant and the PREX-II experiment}",
      journal = {\prd},
         year = 2024,
        month = mar,
       volume = {109},
       number = {6},
          eid = {063017},
        pages = {063017},
          doi = {10.1103/PhysRevD.109.063017},
archivePrefix = {arXiv},
       eprint = {2312.07113},
 primaryClass = {astro-ph.HE},
       adsurl = {https://ui.adsabs.harvard.edu/abs/2024PhRvD.109f3017L}
}

@article{Sagun:2023wit,
       author = {{Sagun}, Violetta and {Giangrandi}, Edoardo and {Dietrich}, Tim and {Ivanytskyi}, Oleksii and {Negreiros}, Rodrigo and {Provid{\^e}ncia}, Constan{\c{c}}a},
        title = "{What Is the Nature of the HESS J1731-347 Compact Object?}",
      journal = {\apj},
         year = 2023,
        month = nov,
       volume = {958},
       number = {1},
          eid = {49},
        pages = {49},
          doi = {10.3847/1538-4357/acfc9e},
archivePrefix = {arXiv},
       eprint = {2306.12326},
 primaryClass = {astro-ph.HE},
       adsurl = {https://ui.adsabs.harvard.edu/abs/2023ApJ...958...49S}
}

@article{Demorest:2010sdm,
    author = "Demorest, Paul and Pennucci, Tim and Ransom, Scott and Roberts, Mallory and Hessels, Jason",
    title = "{Shapiro Delay Measurement of A Two Solar Mass Neutron Star}",
    eprint = "1010.5788",
    archivePrefix = "arXiv",
    primaryClass = "astro-ph.HE",
    doi = "10.1038/nature09466",
    journal = "Nature",
    volume = "467",
    pages = "1081--1083",
    year = "2010"
}

@article{Abbott:2018gmo,
    author = "Abbott, B.P. and others",
    collaboration = "LIGO Scientific, Virgo",
    title = "{GW170817: Measurements of neutron star radii and equation of state}",
    eprint = "1805.11581",
    archivePrefix = "arXiv",
    primaryClass = "gr-qc",
    reportNumber = "LIGO-P1800115",
    doi = "10.1103/PhysRevLett.121.161101",
    journal = "Phys. Rev. Lett.",
    volume = "121",
    number = "16",
    pages = "161101",
    year = "2018"
}

@article{Riley:2019anv,
    author = "Riley, Thomas E. and others",
    title = "{A NICER View of PSR J0030+0451: Millisecond Pulsar Parameter Estimation}",
    eprint = "1912.05702",
    archivePrefix = "arXiv",
    primaryClass = "astro-ph.HE",
    doi = "10.3847/2041-8213/ab481c",
    journal = "Astrophys. J. Lett.",
    volume = "887",
    number = "1",
    pages = "L21",
    year = "2019"
}

@article{Alford:2023dcc,
doi = {10.3847/1538-4357/acaf55},
url = {https://dx.doi.org/10.3847/1538-4357/acaf55},
year = {2023},
month = {feb},
publisher = {The American Astronomical Society},
volume = {944},
number = {1},
pages = {36},
author = {J. A. J. Alford and J. P. Halpern},
title = {Do Central Compact Objects have Carbon Atmospheres?},
journal = {The Astrophysical Journal},
}

@article{Doroshenko:2022nwp,
    author = {Doroshenko, Victor and Suleimanov, Valery and P{\"u}hlhofer, Gerd and Santangelo, Andrea},
    title = "{A strangely light neutron star within a supernova remnant}",
    doi = "10.1038/s41550-022-01800-1",
    journal = "Nature Astron.",
    volume = "6",
    number = "12",
    pages = "1444--1451",
    year = "2022"
}

@article{Alford:2013aca,
    author = "Alford, Mark G. and Han, Sophia and Prakash, Madappa",
    title = "{Generic conditions for stable hybrid stars}",
    eprint = "1302.4732",
    archivePrefix = "arXiv",
    primaryClass = "astro-ph.SR",
    doi = "10.1103/PhysRevD.88.083013",
    journal = "Phys. Rev. D",
    volume = "88",
    number = "8",
    pages = "083013",
    year = "2013"
}

@article{Read:2008iy,
    author = "Read, Jocelyn S. and Lackey, Benjamin D. and Owen, Benjamin J. and Friedman, John L.",
    title = "{Constraints on a phenomenologically parameterized neutron-star equation of state}",
    eprint = "0812.2163",
    archivePrefix = "arXiv",
    primaryClass = "astro-ph",
    doi = "10.1103/PhysRevD.79.124032",
    journal = "Phys. Rev. D",
    volume = "79",
    pages = "124032",
    year = "2009"
}

@ARTICLE{Mariani:2024cas,
       author = {{Mariani}, Mauro and {Ranea-Sandoval}, Ignacio F. and {Lugones}, Germ{\'a}n and {Orsaria}, Milva G.},
        title = "{Could a slow stable hybrid star explain the central compact object in HESS J1731-347?}",
      journal = {\prd},
         year = 2024,
        month = aug,
       volume = {110},
       number = {4},
          eid = {043026},
        pages = {043026},
          doi = {10.1103/PhysRevD.110.043026},
archivePrefix = {arXiv},
       eprint = {2407.06347},
 primaryClass = {astro-ph.HE},
       adsurl = {https://ui.adsabs.harvard.edu/abs/2024PhRvD.110d3026M}
}

@article{Pereira:2017rmp,
    author = "Pereira, Jonas P. and Flores, C\'esar V. and Lugones, Germ\'an",
    title = "{Phase transition effects on the dynamical stability of hybrid neutron stars}",
    eprint = "1706.09371",
    archivePrefix = "arXiv",
    primaryClass = "gr-qc",
    doi = "10.3847/1538-4357/aabfbf",
    journal = "Astrophys. J.",
    volume = "860",
    number = "1",
    pages = "12",
    year = "2018"
}

@article{Baym:1971nsm,
title = {Neutron star matter},
journal = {Nuclear Physics A},
volume = {175},
number = {2},
pages = {225-271},
year = {1971},
issn = {0375-9474},
doi = {https://doi.org/10.1016/0375-9474(71)90281-8},
url = {https://www.sciencedirect.com/science/article/pii/0375947471902818},
author = {Gordon Baym and Hans A. Bethe and Christopher J Pethick}
}

@ARTICLE{Baym:1971tgs,
       author = {{Baym}, Gordon and {Pethick}, Christopher and {Sutherland}, Peter},
        title = "{The Ground State of Matter at High Densities: Equation of State and Stellar Models}",
      journal = {\apj},
         year = 1971,
        month = dec,
       volume = {170},
        pages = {299},
          doi = {10.1086/151216},
       adsurl = {https://ui.adsabs.harvard.edu/abs/1971ApJ...170..299B}
}

@ARTICLE{Antoniadis:2013amp,
   author = {{Antoniadis}, J. and others},
    title = "{A Massive Pulsar in a Compact Relativistic Binary}",
  journal = {Science},
archivePrefix = "arXiv",
   eprint = {1304.6875},
 primaryClass = "astro-ph.HE",
     year = 2013,
    month = apr,
   volume = 340,
    pages = {448},
      doi = {10.1126/science.1233232},
   adsurl = {http://adsabs.harvard.edu/abs/2013Sci...340..448A}
}

@ARTICLE{Cromartie:2020rsd,
       author = {{Cromartie}, H.~T. and others},
        title = "{Relativistic Shapiro delay measurements of an extremely massive millisecond pulsar}",
      journal = {Nature Astronomy},
         year = 2020,
        month = jan,
       volume = {4},
        pages = {72-76},
          doi = {10.1038/s41550-019-0880-2},
archivePrefix = {arXiv},
       eprint = {1904.06759},
 primaryClass = {astro-ph.HE},
       adsurl = {https://ui.adsabs.harvard.edu/abs/2020NatAs...4...72C}
}

@ARTICLE{Fonseca:2021rfa,
       author = {{Fonseca}, E. and others},
        title = "{Refined Mass and Geometric Measurements of the High-mass PSR J0740+6620}",
      journal = {ApJL},
         year = 2021,
        month = jul,
       volume = {915},
       number = {1},
          eid = {L12},
        pages = {L12},
          doi = {10.3847/2041-8213/ac03b8},
archivePrefix = {arXiv},
       eprint = {2104.00880},
 primaryClass = {astro-ph.HE},
       adsurl = {https://ui.adsabs.harvard.edu/abs/2021ApJ...915L..12F}
}

@ARTICLE{Abbott:2017gwa,
   author = {{Abbott}, B.~P. and others},
    title = "{Gravitational Waves and Gamma-Rays from a Binary Neutron Star Merger: GW170817 and GRB 170817A}",
  journal = {\apjl},
archivePrefix = "arXiv",
   eprint = {1710.05834},
 primaryClass = "astro-ph.HE",
     year = 2017,
    month = oct,
   volume = 848,
      eid = {L13},
    pages = {L13},
      doi = {10.3847/2041-8213/aa920c},
   adsurl = {http://adsabs.harvard.edu/abs/2017ApJ...848L..13A}
}

@article{Abbott:2020goo,
       author = {{Abbott}, B.~P. and others},
        title = "{GW190425: Observation of a Compact Binary Coalescence with Total Mass {\ensuremath{\sim}} 3.4 M$_{{\ensuremath{\odot}}}$}",
      journal = {\apjl},
         year = 2020,
        month = mar,
       volume = {892},
       number = {1},
          eid = {L3},
        pages = {L3},
          doi = {10.3847/2041-8213/ab75f5},
archivePrefix = {arXiv},
       eprint = {2001.01761},
 primaryClass = {astro-ph.HE},
       adsurl = {https://ui.adsabs.harvard.edu/abs/2020ApJ...892L...3A}
}

@article{Miller:2019pjm,
	doi = {10.3847/2041-8213/ab50c5},
	year = 2019,
	month = {dec},
	publisher = {American Astronomical Society},
	volume = {887},
	number = {1},
	pages = {L24},
	author = {M. C. Miller and et al.},
	title = {{PSR} J0030+0451 Mass and Radius from {NICER} Data and Implications for the Properties of Neutron Star Matter},
	journal = {Astrophys. J. Lett.}
}

@article{Miller:2021tro,
       author = {{Miller}, M.~C. and others},
        title = "{The Radius of PSR J0740+6620 from NICER and XMM-Newton Data}",
      journal = {\apjl},
         year = 2021,
        month = sep,
       volume = {918},
       number = {2},
          eid = {L28},
        pages = {L28},
          doi = {10.3847/2041-8213/ac089b},
archivePrefix = {arXiv},
       eprint = {2105.06979},
 primaryClass = {astro-ph.HE},
       adsurl = {https://ui.adsabs.harvard.edu/abs/2021ApJ...918L..28M}
}

@ARTICLE{Riley:2021anv,
       author = {{Riley}, Thomas E. and others},
        title = "{A NICER View of the Massive Pulsar PSR J0740+6620 Informed by Radio Timing and XMM-Newton Spectroscopy}",
      journal = {\apjl},
         year = 2021,
        month = sep,
       volume = {918},
       number = {2},
          eid = {L27},
        pages = {L27},
          doi = {10.3847/2041-8213/ac0a81},
archivePrefix = {arXiv},
       eprint = {2105.06980},
 primaryClass = {astro-ph.HE},
       adsurl = {https://ui.adsabs.harvard.edu/abs/2021ApJ...918L..27R}
}

@ARTICLE{Seidov:1971tso,
       author = {{Seidov}, Z.~F.},
        title = "{The Stability of a Star with a Phase Change in General Relativity Theory}",
      journal = {\sovast},
         year = 1971,
        month = oct,
       volume = {15},
        pages = {347},
       url = {https://ui.adsabs.harvard.edu/abs/1971SvA....15..347S}
}

@ARTICLE{kini:2024ctp,
       author = {{Kini}, Yves and others},
        title = "{Constraining the properties of the thermonuclear burst oscillation source XTE J1814-338 through pulse profile modelling}",
      journal = {\mnras},
         year = 2024,
        month = dec,
       volume = {535},
       number = {2},
        pages = {1507-1525},
          doi = {10.1093/mnras/stae2398},
archivePrefix = {arXiv},
       eprint = {2405.10717},
 primaryClass = {astro-ph.HE},
       adsurl = {https://ui.adsabs.harvard.edu/abs/2024MNRAS.535.1507K}
}

@ARTICLE{Annala:2020efq,
       author = {{Annala}, Eemeli and {Gorda}, Tyler and {Kurkela}, Aleksi and {N{\"a}ttil{\"a}}, Joonas and {Vuorinen}, Aleksi},
        title = "{Evidence for quark-matter cores in massive neutron stars}",
      journal = {Nature Physics},
         year = 2020,
        month = jun,
       volume = {16},
       number = {9},
        pages = {907-910},
          doi = {10.1038/s41567-020-0914-9},
archivePrefix = {arXiv},
       eprint = {1903.09121},
 primaryClass = {astro-ph.HE},
       adsurl = {https://ui.adsabs.harvard.edu/abs/2020NatPh..16..907A}
}

@article{Hebeler:2013nza,
    author = "Hebeler, K. and Lattimer, J. M. and Pethick, C. J. and Schwenk, A.",
    title = "{Equation of state and neutron star properties constrained by nuclear physics and observation}",
    eprint = "1303.4662",
    archivePrefix = "arXiv",
    primaryClass = "astro-ph.SR",
    doi = "10.1088/0004-637X/773/1/11",
    journal = "Astrophys. J.",
    volume = "773",
    pages = "11",
    year = "2013"
}

@ARTICLE{OBoyle:2020peo,
       author = {{O'Boyle}, Michael F. and {Markakis}, Charalampos and {Stergioulas}, Nikolaos and {Read}, Jocelyn S.},
        title = "{Parametrized equation of state for neutron star matter with continuous sound speed}",
      journal = {Phys.Rev.D},
         year = 2020,
        month = oct,
       volume = {102},
       number = {8},
          eid = {083027},
        pages = {083027},
          doi = {10.1103/PhysRevD.102.083027},
archivePrefix = {arXiv},
       eprint = {2008.03342},
 primaryClass = {astro-ph.HE},
       adsurl = {https://ui.adsabs.harvard.edu/abs/2020PhRvD.102h3027O}
}

@ARTICLE{Drischler:2020hwd,
       author = {{Drischler}, C. and {Furnstahl}, R.~J. and {Melendez}, J.~A. and {Phillips}, D.~R.},
        title = "{How Well Do We Know the Neutron-Matter Equation of State at the Densities Inside Neutron Stars? A Bayesian Approach with Correlated Uncertainties}",
      journal = {\prl},
         year = 2020,
        month = nov,
       volume = {125},
       number = {20},
          eid = {202702},
        pages = {202702},
          doi = {10.1103/PhysRevLett.125.202702},
archivePrefix = {arXiv},
       eprint = {2004.07232},
 primaryClass = {nucl-th},
       adsurl = {https://ui.adsabs.harvard.edu/abs/2020PhRvL.125t2702D}
}

@article{Drischler:2021lma,
  title = {Limiting masses and radii of neutron stars and their implications},
  author = {Drischler, Christian and Han, Sophia and Lattimer, James M. and Prakash, Madappa and Reddy, Sanjay and Zhao, Tianqi},
  journal = {Phys. Rev. C},
  volume = {103},
  issue = {4},
  pages = {045808},
  numpages = {23},
  year = {2021},
  month = {Apr},
  publisher = {American Physical Society},
  doi = {10.1103/PhysRevC.103.045808},
  url = {https://link.aps.org/doi/10.1103/PhysRevC.103.045808}
}

@article{Shibata:2019ctb,
    author = "Shibata, Masaru and Zhou, Enping and Kiuchi, Kenta and Fujibayashi, Sho",
    title = "{Constraint on the maximum mass of neutron stars using GW170817 event}",
    eprint = "1905.03656",
    archivePrefix = "arXiv",
    primaryClass = "astro-ph.HE",
    doi = "10.1103/PhysRevD.100.023015",
    journal = "Phys. Rev. D",
    volume = "100",
    number = "2",
    pages = "023015",
    year = "2019"
}

@ARTICLE{Rezzolla:2018ugw,
       author = {{Rezzolla}, Luciano and {Most}, Elias R. and {Weih}, Lukas R.},
        title = "{Using Gravitational-wave Observations and Quasi-universal Relations to Constrain the Maximum Mass of Neutron Stars}",
      journal = {\apjl},
         year = 2018,
        month = jan,
       volume = {852},
       number = {2},
          eid = {L25},
        pages = {L25},
          doi = {10.3847/2041-8213/aaa401},
archivePrefix = {arXiv},
       eprint = {1711.00314},
 primaryClass = {astro-ph.HE},
       adsurl = {https://ui.adsabs.harvard.edu/abs/2018ApJ...852L..25R}
}

@ARTICLE{Musolino:2024otm,
       author = {{Musolino}, Carlo and {Ecker}, Christian and {Rezzolla}, Luciano},
        title = "{On the Maximum Mass and Oblateness of Rotating Neutron Stars with Generic Equations of State}",
      journal = {\apj},
         year = 2024,
        month = feb,
       volume = {962},
       number = {1},
          eid = {61},
        pages = {61},
          doi = {10.3847/1538-4357/ad1758},
archivePrefix = {arXiv},
       eprint = {2307.03225},
 primaryClass = {gr-qc},
       adsurl = {https://ui.adsabs.harvard.edu/abs/2024ApJ...962...61M}
}

@ARTICLE{Jarvinen:2022hmo,
       author = {{J{\"a}rvinen}, Matti},
        title = "{Holographic modeling of nuclear matter and neutron stars}",
      journal = {European Physical Journal C},
         year = 2022,
        month = apr,
       volume = {82},
       number = {4},
          eid = {282},
        pages = {282},
          doi = {10.1140/epjc/s10052-022-10227-x},
archivePrefix = {arXiv},
       eprint = {2110.08281},
 primaryClass = {hep-ph},
       adsurl = {https://ui.adsabs.harvard.edu/abs/2022EPJC...82..282J}
}

@ARTICLE{Mariani:2022omh,
       author = {{Mariani}, Mauro and {Tonetto}, Lucas and {Rodr{\'\i}guez}, M. Camila and {Celi}, Marcos O. and {Ranea-Sandoval}, Ignacio F. and {Orsaria}, Milva G. and {P{\'e}rez Mart{\'\i}nez}, Aurora},
        title = "{Oscillating magnetized hybrid stars under the magnifying glass of multimessenger observations}",
      journal = {\mnras},
         year = 2022,
        month = may,
       volume = {512},
       number = {1},
        pages = {517-534},
          doi = {10.1093/mnras/stac546},
archivePrefix = {arXiv},
       eprint = {2202.12222},
 primaryClass = {astro-ph.HE},
       adsurl = {https://ui.adsabs.harvard.edu/abs/2022MNRAS.512..517M}
}

@ARTICLE{Blomqvist:2025sma,
       author = {{Blomqvist}, Sofia and {Ecker}, Christian and {Gorda}, Tyler and {Vuorinen}, Aleksi},
        title = "{Strong model-agnostic constraints for twin-star solutions}",
      journal = {arXiv e-prints},
         year = 2025,
        month = dec,
          eid = {arXiv:2512.19477},
        pages = {arXiv:2512.19477},
          doi = {10.48550/arXiv.2512.19477},
archivePrefix = {arXiv},
       eprint = {2512.19477},
 primaryClass = {astro-ph.HE},
       adsurl = {https://ui.adsabs.harvard.edu/abs/2025arXiv251219477B}
}

@ARTICLE{Komoltsev:2024fop,
       author = {{Komoltsev}, Oleg},
        title = "{First-order phase transitions in the cores of neutron stars}",
      journal = {\prd},
         year = 2024,
        month = oct,
       volume = {110},
       number = {7},
          eid = {L071502},
        pages = {L071502},
          doi = {10.1103/PhysRevD.110.L071502},
archivePrefix = {arXiv},
       eprint = {2404.05637},
 primaryClass = {nucl-th},
       adsurl = {https://ui.adsabs.harvard.edu/abs/2024PhRvD.110g1502K}
}

@Article{Mariani:2026hns,
AUTHOR = {Mariani, Mauro and Ranea-Sandoval, Ignacio F.},
TITLE = {How Neutron Star Observations Point Towards Exotic Matter: Existing Explanations and a Prospective Proposal},
JOURNAL = {Symmetry},
VOLUME = {18},
YEAR = {2026},
NUMBER = {1},
ARTICLE-NUMBER = {27},
URL = {https://www.mdpi.com/2073-8994/18/1/27},
ISSN = {2073-8994},
DOI = {10.3390/sym18010027}
}

@article{Celi:2025etr,
  title = {Exploring the role of ${d}^{*}$ hexaquarks on quark deconfinement and hybrid stars},
  author = {Celi, Marcos O. and Mariani, Mauro and Kumar, Rajesh and Bashkanov, Mikhail and Orsaria, Milva G. and Pastore, Alessandro and Ranea-Sandoval, Ignacio F. and Dexheimer, Veronica},
  journal = {Phys. Rev. D},
  volume = {112},
  issue = {2},
  pages = {023027},
  numpages = {20},
  year = {2025},
  month = {Jul},
  publisher = {American Physical Society},
  doi = {10.1103/3lyv-45jp},
  url = {https://link.aps.org/doi/10.1103/3lyv-45jp}
}

@article{Celi:2025tau,
  title = {Toward a unified hadron-quark equation of state for neutron stars within the relativistic mean-field model},
  author = {Celi, Marcos O. and Mariani, Mauro and Orsaria, Milva G. and Ranea-Sandoval, Ignacio F. and Lugones, Germ\'an},
  journal = {Phys. Rev. D},
  volume = {112},
  issue = {12},
  pages = {123001},
  numpages = {18},
  year = {2025},
  month = {Dec},
  publisher = {American Physical Society},
  doi = {10.1103/ynml-q8zm},
  url = {https://link.aps.org/doi/10.1103/ynml-q8zm}
}

@article{Lenzi:2023hsw,
  title = {Hybrid stars with reactive interfaces: Analysis within the Nambu--Jona-Lasinio model},
  author = {Lenzi, C. H. and Lugones, G. and Vasquez, C.},
  journal = {Phys. Rev. D},
  volume = {107},
  issue = {8},
  pages = {083025},
  numpages = {12},
  year = {2023},
  month = {Apr},
  publisher = {American Physical Society},
  doi = {10.1103/PhysRevD.107.083025},
  url = {https://link.aps.org/doi/10.1103/PhysRevD.107.083025}
}

@article{ranea:2017csi,
  title = {Color superconductivity in compact stellar hybrid configurations},
  author = {Ranea-Sandoval, Ignacio F. and Orsaria, Milva G. and Han, Sophia and Weber, Fridolin and Spinella, William M.},
  journal = {Phys. Rev. C},
  volume = {96},
  issue = {6},
  pages = {065807},
  numpages = {13},
  year = {2017},
  month = {Dec},
  publisher = {American Physical Society},
  doi = {10.1103/PhysRevC.96.065807},
  url = {https://link.aps.org/doi/10.1103/PhysRevC.96.065807}
}

@ARTICLE{ranea:2016css,
       author = {{Ranea-Sandoval}, Ignacio F. and {Han}, Sophia and {Orsaria}, Milva G. and {Contrera}, Gustavo A. and {Weber}, Fridolin and {Alford}, Mark G.},
        title = "{Constant-sound-speed parametrization for Nambu-Jona-Lasinio models of quark matter in hybrid stars}",
      journal = {\prc},
         year = 2016,
        month = apr,
       volume = {93},
       number = {4},
          eid = {045812},
        pages = {045812},
          doi = {10.1103/PhysRevC.93.045812},
archivePrefix = {arXiv},
       eprint = {1512.09183},
 primaryClass = {nucl-th},
       adsurl = {https://ui.adsabs.harvard.edu/abs/2016PhRvC..93d5812R}
}

@article{tonetto:2020dgm,
       author = {{Tonetto}, L. and {Lugones}, G.},
        title = "{Discontinuity gravity modes in hybrid stars: Assessing the role of rapid and slow phase conversions}",
      journal = {\prd},
         year = 2020,
        month = jun,
       volume = {101},
       number = {12},
          eid = {123029},
        pages = {123029},
          doi = {10.1103/PhysRevD.101.123029},
archivePrefix = {arXiv},
       eprint = {2003.01259},
 primaryClass = {astro-ph.HE},
       adsurl = {https://ui.adsabs.harvard.edu/abs/2020PhRvD.101l3029T}
}

@ARTICLE{Salmi:2024anv,
       author = {{Salmi}, Tuomo and others},
        title = "{A NICER View of PSR J1231‑1411: A Complex Case}",
      journal = {\apj},
         year = 2024,
        month = nov,
       volume = {976},
       number = {1},
          eid = {58},
        pages = {58},
          doi = {10.3847/1538-4357/ad81d2},
archivePrefix = {arXiv},
       eprint = {2409.14923},
 primaryClass = {astro-ph.HE},
       adsurl = {https://ui.adsabs.harvard.edu/abs/2024ApJ...976...58S}
}

@ARTICLE{Choudhury2024anv,
       author = {{Choudhury}, Devarshi and others},
        title = "{A NICER View of the Nearest and Brightest Millisecond Pulsar: PSR J0437{\textendash}4715}",
      journal = {\apjl},
         year = 2024,
        month = aug,
       volume = {971},
       number = {1},
          eid = {L20},
        pages = {L20},
          doi = {10.3847/2041-8213/ad5a6f},
archivePrefix = {arXiv},
       eprint = {2407.06789},
 primaryClass = {astro-ph.HE},
       adsurl = {https://ui.adsabs.harvard.edu/abs/2024ApJ...971L..20C}
}

@ARTICLE{Mauviard:2025anv,
       author = {{Mauviard}, Lucien and others},
        title = "{A NICER View of the 1.4 M$_{{\ensuremath{\odot}}}$ Edge-on Pulsar PSR J0614-3329}",
      journal = {\apj},
         year = 2025,
        month = dec,
       volume = {995},
       number = {1},
          eid = {60},
        pages = {60},
          doi = {10.3847/1538-4357/ae145d},
archivePrefix = {arXiv},
       eprint = {2506.14883},
 primaryClass = {astro-ph.HE},
       adsurl = {https://ui.adsabs.harvard.edu/abs/2025ApJ...995...60M}
}

@article{Holt:2017eos,
  title = {Equation of state of nuclear and neutron matter at third-order in perturbation theory from chiral effective field theory},
  author = {Holt, J. W. and Kaiser, N.},
  journal = {Phys. Rev. C},
  volume = {95},
  issue = {3},
  pages = {034326},
  numpages = {9},
  year = {2017},
  month = {Mar},
  publisher = {American Physical Society},
  doi = {10.1103/PhysRevC.95.034326},
  url = {https://link.aps.org/doi/10.1103/PhysRevC.95.034326}
}

@article{Hu:2017nmp,
  title = {Nuclear matter properties with nucleon-nucleon forces up to fifth order in the chiral expansion},
  author = {Hu, Jinniu and Zhang, Ying and Epelbaum, Evgeny and Mei\ss{}ner, Ulf-G. and Meng, Jie},
  journal = {Phys. Rev. C},
  volume = {96},
  issue = {3},
  pages = {034307},
  numpages = {7},
  year = {2017},
  month = {Sep},
  publisher = {American Physical Society},
  doi = {10.1103/PhysRevC.96.034307},
  url = {https://link.aps.org/doi/10.1103/PhysRevC.96.034307}
}

@article{Lynn:2016ctn,
  title = {Chiral Three-Nucleon Interactions in Light Nuclei, Neutron-$\ensuremath{\alpha}$ Scattering, and Neutron Matter},
  author = {Lynn, J. E. and Tews, I. and Carlson, J. and Gandolfi, S. and Gezerlis, A. and Schmidt, K. E. and Schwenk, A.},
  journal = {Phys. Rev. Lett.},
  volume = {116},
  issue = {6},
  pages = {062501},
  numpages = {5},
  year = {2016},
  month = {Feb},
  publisher = {American Physical Society},
  doi = {10.1103/PhysRevLett.116.062501},
  url = {https://link.aps.org/doi/10.1103/PhysRevLett.116.062501}
}

@article{Gorda:2018gpy,
    author = {Gorda, Tyler and Kurkela, Aleksi and Romatschke, Paul and S\"appi, Matias and Vuorinen, Aleksi},
    title = "{Next-to-Next-to-Next-to-Leading Order Pressure of Cold Quark Matter: Leading Logarithm}",
    eprint = "1807.04120",
    archivePrefix = "arXiv",
    primaryClass = "hep-ph",
    reportNumber = "CERN-TH-2018-230, HIP-2018-13/TH",
    doi = "10.1103/PhysRevLett.121.202701",
    journal = "Phys. Rev. Lett.",
    volume = "121",
    number = "20",
    pages = "202701",
    year = "2018"
}

@ARTICLE{Mariani:2024csi,
       author = {{Mariani}, Mauro and {Albertus}, Conrado and {Alessandroni}, M. del Rosario and {Orsaria}, Milva G. and {P{\'e}rez-Garc{\'\i}a}, M. {\'A}ngeles and {Ranea-Sandoval}, Ignacio F.},
        title = "{Constraining self-interacting fermionic dark matter in admixed neutron stars using multimessenger astronomy}",
      journal = {\mnras},
         year = 2024,
        month = jan,
       volume = {527},
       number = {3},
        pages = {6795-6806},
          doi = {10.1093/mnras/stad3658},
archivePrefix = {arXiv},
       eprint = {2311.14004},
 primaryClass = {astro-ph.HE},
       adsurl = {https://ui.adsabs.harvard.edu/abs/2024MNRAS.527.6795M}
}

@ARTICLE{Drago:2016tso1,
       author = {{Drago}, Alessandro and {Lavagno}, Andrea and {Pagliara}, Giuseppe and {Pigato}, Daniele},
        title = "{The scenario of two families of compact stars. Part 1. Equations of state, mass-radius relations and binary systems}",
      journal = {European Physical Journal A},
         year = 2016,
        month = feb,
       volume = {52},
          eid = {40},
        pages = {40},
          doi = {10.1140/epja/i2016-16040-3},
       adsurl = {https://ui.adsabs.harvard.edu/abs/2016EPJA...52...40D}
}

\end{document}